\documentclass[12pt, fullpage, a4paper]{article}
\usepackage{graphicx, latexsym}
\usepackage{setspace}
\usepackage{subfigure}
\usepackage{apacite}
\usepackage{amssymb, amsmath, amsthm}
\usepackage{bm}
\usepackage{epstopdf}
\usepackage{rotating} %used for \sidewaystable
\usepackage{apacite}
\usepackage{booktabs} %used for \toprule
\usepackage{multirow} %used for \multirow and \multicolumn
\usepackage{pifont}
\usepackage{listings}
\usepackage[svgnames]{xcolor}
\begin{document}
\title{Graphical and numerical diagnostic tools to assess multiple imputation models by posterior predictive checking}
\author{Mingyang Cai, Stef van Buuren and Gerko Vink}
\date{}
\maketitle
\begin{abstract}
Missing data are often dealt with multiple imputation. A crucial part of the multiple imputation process is selecting sensible models to generate plausible values for the incomplete data. A method based on posterior predictive checking is proposed to diagnose imputation models based on posterior predictive checking. To assess the congeniality of imputation models, the proposed diagnostic method compares the observed data with their replicates generated under corresponding posterior predictive distributions. If the imputation model is congenial with the substantive model, the observed data are expected to be located in the centre of corresponding predictive posterior distributions. Simulation and application are designed to investigate the proposed diagnostic method for parametric and semi-parametric imputation approaches, continuous and discrete incomplete variables, univariate and multivariate missingness patterns. The results show the validity of the proposed diagnostic method.  	
\end{abstract}

\section{Introduction}
Multiple imputation (MI) is a popular approach for the analysis of incomplete datasets. It involves generating several plausible imputed datasets and aggregating different results into a single inference. Missing cells are filled with synthetic data drawn from corresponding posterior predictive distributions. This procedure is repeated multiple times, resulting in several imputed datasets. The parameters of scientific interest are then estimated for each imputed dataset by complete-data analyses. Finally, different estimates are pooled into one inference using Rubin's rule, which accounts for within and across imputation uncertainty \cite{RubinD1987}.  

The validity of post-imputation analyses relies on the congeniality of the imputation model and the analysis model (Meng, 1994\nocite{meng1994multiple}; Xie \& Meng, 2017\nocite{xie2017dissecting}; Bartlett\& Hughes, 2020\nocite{bartlett2020bootstrap}). If the imputation and analysis models are congenial and correctly specified, Rubin’s rules will give consistent estimates \cite{RubinD1987}. Hence, a crucial part of the multiple imputation process is selecting sensible models to generate plausible values for the incomplete data. However, model selection is not a trivial process in practice since there can be a wide array of candidate models to check. Therefore, researchers should consider which variables, interaction terms, and nonlinear terms are included based on the scientific interest and data features. 

Despite the popularity of multiple imputation, there are only a few imputation model diagnostic methodologies. One standard diagnostic method is to compare distributions of the observed with imputed data (Van Buuren, 2018;\nocite{Buuren2018} Abayomi et al., 2008\nocite{abayomi2008diagnostics}). Plausible imputation models would generate imputed values that have a similar distribution to the observed data. Although missing at random (MAR) mechanisms would also induce the discrepancies between the observed and imputed data, any dramatic departures that the observed data features cannot explain are evidence of potential imputation model misspecification. Reliable interpretation of the observed and imputed data's discrepancies could be derived from external knowledge about the incomplete variables and the missingness mechanisms (Abayomi et al., 2008).

The idea to evaluate the validity of scientific models with multiply imputed data is not new. Bondarenko and Raghunathan (2016)\nocite{bondarenko2016graphical} proposed an advanced diagnostic method to compare the distributions of the observed with imputed data conditional on the probability of missingness. Gelman et al. (1998)\nocite{gelman1998not} applied cross-validation to check the fit of a hierarchical Bayesian model in the study of 51 public opinion polls preceding the 1988 U.S. Presidential election. Gelman et al. (2005)\nocite{gelman2005multiple} also proposed to apply graphical posterior predictive inference on the test statistics for model checking with missing and latent data. If regression-based imputation approaches are involved, the conventional regression diagnostics, such as plots of residuals and outliers, are helpful (White et al., 2011)\nocite{white2011multiple}. A comprehensive overview of model diagnostic in multiple imputation is available in Nguyen et al. (2017)\nocite{nguyen2017model}.

Posterior predictive checking (PPC) has been proposed as an alternative method for the imputation model diagnostic (Gelman et al., 2005; He and Zaslavsky, 2012; Nguyen et al., 2015\nocite{nguyen2015posterior}). PPC is a Bayesian model checking approach that compares the replicated data drawn from the corresponding posterior predictive distribution to the observed data. If the model lacks fit, there could be a discrepancy between the replicated and observed data.

He \& Zaslavsky (2012) and Nguyen et al. (2015) applied posterior predictive checking to assess the inadequacies of the joint imputation model with one or more test quantities relevant to the scientific interest. To evaluate the `usability' of imputation models with respect to the test statistics, analysts compare the estimates for the complete data to their replicates. Comparisons of the complete data and its replicates ensure the calculation of test quantities with general missingness patterns. However, it also results in sensitivity to the amount of missingness.

This manuscript proposes and evaluates the implementation of posterior predictive checking for imputation techniques. The general idea is that if the imputation model is congenial to the substantive model, the expected value of the data (whether observed or missing) is in the centre of corresponding predictive posterior distributions. We compare the observed data to their posterior replicates generated under the imputation model and evaluate the posterior distributions of all observed data points. This distinguishes our approach from the posterior predictive checking of imputation models by applying target analyses. We demonstrate:
\begin{enumerate}
	\item that PPC can be generalised to variable-by-variable imputation techniques; 
	\item that PPC can be used to identify the imputation model that conforms most to the true data generating model;
	\item that PPC can be used as a model evaluation technique to identify the better substantive analysis model;
	\item how to perform PPC with \texttt{MICE} in \texttt{R} on a real-life data set (Van Buuren and Groothuis-Oudshoorn, 2011\nocite{van2011mice});
	\item that this PPC approach is not sensitive to the amount of missing data.
\end{enumerate}
The remainder of this manuscript is organised as follows. In section 2, we review the posterior predictive checking of the imputation model by applying the target analysis proposed by He \& Zaslavsky (2012)\nocite{he2012diagnosing}. In section 3, we provide an overview of the \texttt{MICE} package and the underlying imputation algorithm: fully conditional specification (FCS). We also further point out the necessity of extending the posterior predictive checking of the imputation model so that the diagnostics would apply to the \texttt{MICE} algorithm. In section 4, we evaluate the performance of the proposed diagnostic approach with simulation studies. In section 5, we show the results of simulation studies. In section 6, we apply the proposed diagnostic approach to the body mass index (BMI) data of Dutch adults. In section 7, we conclude with a discussion of our findings.

\section{Posterior predictive checking (PPC)}
\subsection{Posterior predictive checking}
Without incomplete variables, PPC compares the observed data $y$ with the replicated data $y^{rep}$ which are simulated from the posterior predictive distribution, with parameter $\theta$:
\begin{equation}
	\begin{array}{ll}
		p(y^{rep}|y) = \int p(y^{rep}|\theta)p(\theta|y)d\theta
	\end{array} 
\end{equation}
To detect the discrepancy between the model and the data, we define test quantities $T$ that connect the model diagnostics to the scientific interests and estimate them for both observed and replicated data. For example, if the substantive analysis is a regression analysis, the test quantities could be the regression coefficients. Misfits of the model with respect to the data could be summarised by the posterior predictive p-value, which is the probability that the replicated data are more extreme than the observed data, with respect to the selected test quantities $T$ \cite{gelman2013bayesian}:
\begin{equation}
	\begin{array}{ll}
		p_{B} &= Pr(T(y^{rep}, \theta) \ge T(y, \theta)|y)\\
		&= \int\int I_{T(y^{rep}, \theta) \ge T(y, \theta)}p(y^{rep}|\theta)p(\theta|y)dy^{rep}d\theta,
	\end{array} 
\end{equation}
where $I$ is the indicator function. An extreme p-value (close to 0 or 1) implies the suspicion on the fit of the model since a consistent discrepancy between test quantities $T(y^{rep}, \theta)$ and $T(y, \theta)$ cannot be explained by the simulation variance. 

Posterior predictive checking has been widely used for model diagnostics in applied Bayesian analysis (Gelman et al., 2013, chapter 6), and the posterior predictive distribution is usually calculated by simulation. Suppose we have $N$ draws of model parameters from its posterior distribution $\theta_j, j=1,\dots,N$, we then generate a replicated data for every theta $\theta_j$. The PPC compares test quantities based on observed data with the empirical predictive distribution of test quantities. The estimated posterior predictive p-value is the proportion of these \texttt{N} simulations for which $T_{j}(y^{rep}, \theta) > T_{j}(y, \theta)$. It is noticeable that PPC's application for the imputation model diagnostic is not based on the hypothesis test perspective. Hence, there is no underlying assumed distribution for the posterior predictive p-value in this case. The posterior predictive p-value indicates whether the model would provide plausible inference based on the data with respect to the selected test quantities.

To perform multiple imputation model checking with PPC, we compare the completed data, the combination of the observed and imputed data, with its replications. Gelman et al. (2005) applied graphical PPC to visualise test quantities comparisons based on completed and replicated data. He \& Zaslavsky (2012) and Nguyen et al. (2015) developed numerical posterior predictive checks as target analyses to the joint imputation model. He and Zaslavsky (2012) proposed two kinds of discrepancies, completed data discrepancy and expected completed-data discrepancy, and the approaches to calculate corresponding posterior predictive p-values. We briefly introduce these discrepancies and p-values for the completeness of PPC for MI models.

\subsection{Complete data discrepancy}
When there are incomplete variables, the hypothetically complete data $y_{com}$ consists of the observed data $y_{obs}$ and the missing data $y_{mis}$ ($y_{com} = (y_{obs}, y_{mis})$). To assess the completed-data discrepancy $T(y_{com}^{rep}, \theta) - T(y_{com}, \theta)$, we draw imputed values for incomplete variables $y_{mis}$ and the replication of the complete data $y_{com}^{rep}$ from their posterior predictive distribution:
\begin{equation}
	\begin{array}{ll}
		p(y_{com}^{rep}, y_{mis}|y_{obs}) = \int p(y_{com}^{rep}|\theta)p(y_{mis}|y_{obs}, \theta)p(\theta|y_{obs})d\theta.
	\end{array} 
\end{equation}
To assess the model fit, we calculate the posterior predictive p value as :
\begin{equation}
	\begin{array}{ll}
		p_{B, com} &= Pr(T(y_{com}^{rep}) \ge T(y_{com})|y_{obs})\\
		&= \int\int I_{T(y_{com}^{rep}) \ge T(y_{com})}p(y_{com}^{rep}, y_{mis}|y_{obs})dy_{com}^{rep}dy_{mis}
	\end{array} 
\end{equation}
The simulation process to estimate p-value proposed by He and Zaslavsky (2012) is:
\begin{enumerate}
	\item Simulate $N$ draws of $\theta$ from the corresponding posterior distribution $p(\theta|y_{obs})$
	\item For each $\theta_{j}, j=1, \dots, N$, impute $y_{mis}^j$ from $p(y_{mis}|y_{obs}, \theta_{j})$ and simulate the replicated data $y_{com}^{rep, j}$ from $p(y_{com}^{rep}|\theta_{j})$
\end{enumerate}
A $p_{B, com}$, which is close to 0 or 1, implies the discrepancy between the model and the data with respect to the selected test quantities.

\subsection{Expected complete data discrepancy}
He and Zaslavsky (2012) noticed that the power of completed-data discrepancy is weakened because the variance of imputed data across complete data $y_{com}$ and replicated data $y_{com}^{rep}$ increase the variance of the test quantities. He and Zaslavsky (2012) reduced the variance of completed-data discrepancy by calculating the expectation value of missing data for each model parameter draw. 
The modification of p-value $p_{B, ecom}$ would be:  
\begin{equation}
	\begin{array}{ll}
		p_{B, ecom} &= Pr(E[T(y_{com}^{rep})|y_{obs}^{rep}, y_{obs}] \ge E[T(y_{com})|y_{obs}^{rep}, y_{obs}]|y_{obs})\\
		&= \int\int I_{E[T(y_{com}^{rep})|y_{obs}^{rep}, y_{obs}] \ge E[T(y_{com})|y_{obs}^{rep}, y_{obs}]}p(y_{obs}^{rep}, y_{obs})dy_{obs}^{rep}, 
	\end{array} 
\end{equation}
where $E$ is the notation of the expected value. 

Again, the nested simulation process to calculate the p-value $p_{B, ecom}$ is:
\begin{enumerate}
	\item Simulate $N_{1}$ draws of $\theta$ from the corresponding posterior distribution $p(\theta|y_{obs})$
	\item For each $\theta_{j}, j=1, \dots, N_{1}$, impute $y_{mis}^j$ from $p(y_{mis}|y_{obs}, \theta_{j})$ and simulate the replicated data $y_{com}^{rep, j}$ from $p(y_{com}^{rep}|\theta_{j})$
	\item For each j-th replicate, calculate the mean discrepancy by setting $y_{mis}^j$ and $y_{com}^{rep, j}$ to missing and overimputing them with the same paramters $\theta_{j}$ over $N_{2}$ draws $y_{mis}^{j, k}$ and $y_{com}^{rep, j, k}, k = 1, \dots, N_{2}$. Calculate the difference : $D_{j, k} = T(y_{obs}^{rep, j}, y_{mis}^{rep, j, k}) - T(y_{obs}, y_{mis}^{rep, j, k})$ over $N_{2}$ draws and then average the difference for the j-th replicate : $\bar{D}_{j.} = \sum_{1}^{k}D_{j, k}/k$
	\item Calculate $p_{B, ecom}$ as the proportion of these $N_{1}$ estimates that are positive, $\bar{D}_{j.} \ge 0$    
\end{enumerate}

He and Zaslavsky (2012) evaluated whether the PPC could detect the uncongeniality of the imputation model. Nguyen et al. (2015) investigated the performance of PPC in other imputation model misspecification scenarios, such as ignoring the response variable and auxiliary variables or failing to transform skewed variables. The PPC approach proposed by He and Zaslavsky (2012) is based on the joint imputation model. The imputation model for diagnostic is a joint distribution for the observed data, and the test quantities depend on multiple variables and parameters.

\section{\texttt{MICE} package}
Fully conditional specification (FCS) is a popular approach for multiple imputation. It attempts to specify an imputation model for each incomplete variable $Y_j, j = 1, \dots, p$ conditional on all the other variables $P(Y_j | Y_{-j}, \theta_{j})$, with parameter $\theta_{j}$. It generates imputations iteratively over all incomplete variables after an initial imputation, such as mean imputation or random draw from observed values. Let $Y_{j}^{t} = (Y_{j}^{obs}, Y_{j}^{mis(t)})$ denote the observed and imputed values of variable $Y_{j}$ at iteration $t$ and $Y_{-j}^{t} = (Y_{1}^{t}, \dots, Y_{j-1}^{t}, Y_{j+1}^{t-1}, \dots, Y_{p}^{t-1})$. Given the most recent imputations of the other incomplete variables $Y_{j}^{t}$ at iteration $t$, the algorithm of generating imputations for the incomplete variable $Y_{j}$ consists of the following draws:
\begin{align*}
	\theta_{j}^{t} \sim f(\theta_{j})f(Y_{j}^{obs}|Y_{-j}^{t}, \theta_{j})\\
	Y_{j}^{mis(t)} \sim f(Y_{j}^{mis}|Y_{-j}^{t}, \theta_{j}^{t}),
\end{align*}
where $f(\theta_{j})$ is the prior distribution for the parameter of the imputation model $\theta_{j}$.
The FCS is an attractive imputation approach because of its flexibility in imputation model specification. It is known under different names: chained equations stochastic relaxation, variable-by-variable imputation, switching regression, sequential regressions, ordered pseudo-Gibbs sampler, partially incompatible MCMC, and iterated univariate imputation \cite{van2007multiple}.

Multivariate Imputation by Chained Equations (\texttt{MICE}) is the name of software for imputing incomplete multivariate data by Fully Conditional Specification. It has developed into the de facto standard for imputation in \texttt{R} and is increasingly being adopted in Python (e.g., statsmodels (imputer function) \& miceforest). The \texttt{MICE} package creates functions for three components of FCS: imputation, analysis, and pooling. 
\begin{figure}[ht!]
	\centering
	\includegraphics[scale=.2]{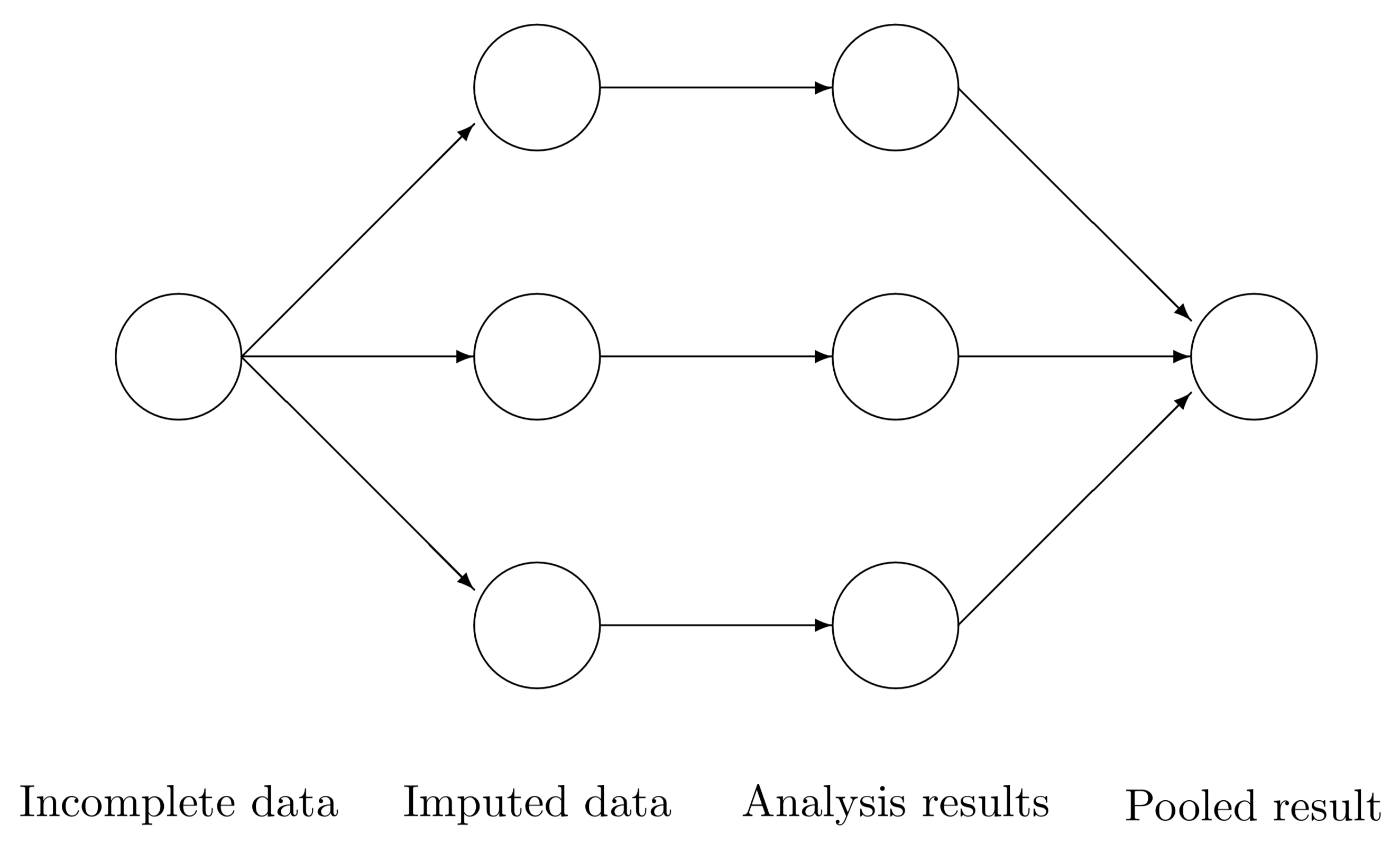}
	\caption{Main steps used in \texttt{MICE} (Van Buuren \& Groothuis-Oudshoorn, 2011)}
	\label{fig6_1}
\end{figure} 
Figure \ref{fig6_1} illustrates how MICE solves a missing data problem by generating 3 imputed datasets. Three imputed datasets are generated with function \textbf{mice()}. Analysis are performed on every imputed dataset by \textbf{with()} function and combined into a single inference with function \textbf{pool()}. The software stores the output of each step in a particular class: \textbf{mids}, \textbf{mira} and \textbf{mipo}. More details about \texttt{MICE} package can be found in van Buuren \& Groothuis-Oudshoorn (2011).\nocite{van2011mice}

Two features of the software motivate our research. First, the default imputation method for numerical missing data is predictive mean matching (PMM) (Little, 1988)\nocite{Little1988}.
\begin{lstlisting}
	library(mice, warn.conflicts = FALSE)
	imp <- mice(nhanes, print = FALSE)
	imp$method
\end{lstlisting}
\begin{verbatim}
	##   age   bmi   hyp   chl 
	##    "" "pmm" "pmm" "pmm"
\end{verbatim}
It generates imputations for a missing cell from its $p$ nearest points. Predictive mean matching is a semi-parametric imputation approach that is proven to perform well in a wide range of scenarios (De Waal et al., 2011;\nocite{de2011handbook} Siddique and Belin, 2007;\nocite{siddique2008multiple} Su et al., 2011;\nocite{su2011multiple} Van Buuren, 2018; Van Buuren and Groothuis-Oudshoorn, 2011; Vink et al., 2014;\nocite{vink2014predictive} Vink et al., 2015;\nocite{vink2015partioned} White et al., 2011; Yu, Burton and Rivero-Arias, 2007\nocite{yu2007evaluation}). The attractive advantage of PMM is that the imputations fall consistently within the range of the observed sample space (Heeringa et al., 2001;\nocite{heeringa2001multivariate} Van Buuren, 2018; Vink et al., 2014; Vink et al., 2015; White et al., 2011; Yu et al., 2007). For instance, PMM prevents imputing negative values for data that are strictly non-negative. Second, \textbf{mids} only stores imputed datasets not the estimated parameters of the imputation models (Hoogland et al., 2020\nocite{hoogland2020handling}).

Based on the features of \texttt{MICE} package discussed above, it is necessary to investigate whether PPC could check the donor selection procedure of PMM and perform PPC based on the observed data itself instead of the target statistics. He and Zaslavsky (2012) briefly discussed the approach to checking imputation models for subsets of incomplete variables. However, they assumed that the imputations of the remaining variables (excluding the incomplete variables of interest in an assessment) are adequate. Therefore, we also evaluate the performance of PPC when relaxing this assumption in the application section.

The implementation of PPC in \texttt{MICE} (version 3.13.15) is straightforward. A new argument \texttt{where} is included in \texttt{mice} function which allows us to replace the observed data by randomly drawing values from the predictive posterior distribution \cite{volker2021anonymiced}. Here is an example to generate replications of the observed data.   
\begin{lstlisting}
	to_imp <- as.data.frame(!is.na(nhanes)) 
	imp <- mice(nhanes, where = to_imp, print = FALSE)
\end{lstlisting}

The observed data pattern may not be conducive to calculating test quantities involving multiple incomplete variables. Hence, the complete data discrepancy discussed in section 2.2 calculates test quantities based on the imputed data. In such a case, the diagnosis is sensitive to the amount of missing data. However, we propose to calculate the discrepancy based on the observed data. The proposed method compares the observed data to the corresponding predictive posterior distributions generated over multiple imputations. If the imputed model fits the observed data, the observed data will appear like a random draw from the corresponding posterior distribution.

\section{Simulation Study}
We carried out a simulation study to investigate the performance of the proposed diagnostic approach. For illustrative purposes, the simulation study consisted of diagnostics under three analysis models: 1) a quadratic equation with an incomplete outcome 2) a quadratic equation with incomplete covariates, and 3) a generalised linear model with an incomplete binary outcome. The proposed diagnostic method natually could be applied to more generalized analysis models. 

All these scenarios are designed with several factors including missingness proportion (30\%, 50\%, 80\%), missingness mechanisms (a missing completely at random (MCAR) and a right-tailed missing at random (MARr) mechanism), nominal levels of the confidence interval (75\%, 95\%) and different imputation models. We evaluated whether the proposed diagnostic method could identify the congenial imputation model for continuous and discrete incomplete variables under the first and the third scenarios. We also investigated the performance of the proposed diagnostic method on the donor selection procedure of predictive mean matching under the second scenario. The sample size and the number of imputed datasets were set to be 1000 and 50 separately in all simulations.

We induced missingness with the \texttt{ampute()} function in the simulation study. Generally, \texttt{ampute()} is a convenient function in \texttt{MICE} package to generate missing data for simulation purposes (Schouten et al. 2018)\nocite{Schouten2018}. We considered missing completely at random (MCAR) mechanism where the probability of missingness is equal for every cell as well as right-tailed missing at random (MARr) mechanism where higher values of covariates have a higher probability of being unobserved. In the algorithm of \texttt{ampute()} function, the probability of missingness is allocated with different logistic functions of the weighted sum score (\emph{wss}), which is a linear combination of covariates correlated with the probability of missingness:
\begin{equation}
	\begin{array}{ll}
		wss_{i} = w_{i}x_{1i} + w_{i}x_{2i} + \dots + w_{i}x_{mi}
	\end{array} 
\end{equation}
The weight $w_i$ is pre-specified to reflect the influence of the variable $x_{i}$ on the probability of missingness. For instance, if the formation of a weighted sum score is:
$wss = x_1 + x_2$, the probability of missingness is determined by both $x_1$ and $x_2$ with the equal effects. It is noticable that the influence of the weights is relative. $wss = 2x_1 + 2x_2$ will have the same effct as $wss = x_1 + x_2$. More specifically, under MARr mechanism, candidates with higher values of weighted sum score have a higher probability of being unobserved when applying the \texttt{ampute()} function to generate missing data.

\subsection{Quadratic equation with an incomplete outcome}
In the first simulation study, we considered a partially observed variable $Y$ and a fully observed variable $X$. The data was generated from : $X \sim \text{uni}(-3, 3)$, $Y = X + X^2 + \epsilon$, where $\epsilon \sim \mathcal{N}(0, 1)$. The scientific model was indeed a quadratic model. Under MARr mechanism, the probability of missingness for the incomplete variable $Y$ was completely determined by variable $X$ ($wss_{Y} = X$). We considered two imputation models for the incomplete response $Y$: one is a linear regression of $Y$ on $X$, and the other is a quadratic regression of $Y$ on $X$. The linear regression impuation model is expected to be uncongenial and have a worse performance.

\subsection{Quadratic equation with incomplete covariates}
In the second simulation study, the response variable $Y$ was generated from a normal distribution: $Y = X + X^2 + \epsilon$, where $\epsilon \sim \mathcal{N}(0, 1)$ and the covariate $X$ followed a standard normal distribution. In this simulation study, the response variable $Y$ was completely observed while the covariate $X$ and the corresponding quadratic term $X^2$ were jointly missing for a fraction of the cases. There were no cases with missing cells on either $X$ or $X^2$. Under MARr mechanism, the probability of missingness for jointly missingness of ${X, X^2}$ was completely determined by variable $Y$ ($wss_{X, X^2} = Y$). We compared two semi-parametric methods, predictive mean matching (PMM) and polynomial combination (PC) with a parametric method, the substantive model compatible fully conditional specification (SMC-FCS) (Vink and van Buuren, 2013, Bartlett et al., 2015, Cai and Vink, 2022\nocite{cai2022note}). The PC and SMC-FCS methods are two accepted methods to impute linear regression with quadratic terms. The PC method is an extension of PMM but applies a different donor selection procedure. We expect that PC and SMC-FCS outperform PMM.

\subsection{Generalized linear model for discrete variables}
The final simulation study considered a partially observed binary $Y$ and two complete covariates $X$ and $Z$. The model of scientific interest was : $Pr (Y = 1 | X, Z) = \text{exp}(X + Z) / 1 + \text{exp}(X + Z)$,
where $x \sim \text{uni}(-3 , 3)$ and $Z \sim \mathcal{N}(1, 1)$. Under MARr mechanism, the weights of variables $X$ and $Z$ in determining the probability of missingness for $Y$ were set to be equal ($wss_{Y} = X + Z$). Since the logistic regression actually models the probability of assignment, we investigated the plot of deviance and calculated the sum of squared deviance divided by the sample size. There were two candidate models: a logistic regression of $Y$ on $X$ and $Z$ and a logistic regression of $Y$ on $Z$ only. The imputation model including both predictors $X$ and $Z$ is expected to provide more sensible imputations.

\section{Simulation results}
In this section, we present the simulation results of the proposed diagnostic method under three different scenarios. We construct the nominal confidence interval for each observed data point based on the empirical distribution generated by the corresponding multiple imputed values. For numerical assessment, we estimated the rate of coverage by which the nominal confidence intervals covers the observed data points (COV), the mean of the distance between the observed data and the mean of corresponding predictive posterior distributions (Distance), and the average width of the confidence intervals (CIW). Since the incomplete variable $Y$ in section 4.1 is the conditionally normal distribution and the incomplete variable $X$ in section 4.2 is the normal distribution, the mean of corresponding predictive posterior distributions is a valid representation of the centre of corresponding posterior distributions. In such a case, the selected quantities (COV, Distance and CIW) could be used to inspect the discrepancy between observed and replicated data. We expect the better-fitted imputation model to derive a smaller Distance and CIW. 

We also provided graphical analyses with scatterplots, density plots, and distribution plots, which show observed values, upper and lower bounds of confidence intervals for each observed data point. The proposed diagnostic approach is performed on a variable-by-variable basis. Sometimes a single plot or summarised statistic is inadequate to arrive at a conclusion. Conducting PPC with various tools would provide a more comprehensive evaluation of the imputation model.

\subsection{Quadratic equation with an incomplete outcome}
Table \ref{tab6_1} shows the results of the simulation study when the substantive model is a quadratic equation with an incomplete outcome. Since we only generated one incomplete dataset and repeated imputing it 50 times, all coverage rates were close to the pre-specified nominal level. However, when the imputation model was misspecified as a linear regression model, the average distance was larger than the average distance under the correct specification of the imputation model (linear regression with a quadratic term). It conforms to our intuitive idea that the data would be close to the centre of predictive posterior distributions if the model fits. The variance of the incomplete variable $Y$ was set as 1, which implied that the width of 95\% nominal confidence interval is approximate 3.92 (1.96 $\times$ 2) and the width of 75\% nominal confidence interval is approximate 2.3 (1.15 $\times$ 2). When the imputation model was correctly specified, the estimated average width of the confidence interval was unbiased. However, the variance of $Y$ was overestimated when the imputation model was linear.

\begin{figure}[b]
	\begin{center}
		\resizebox{\textwidth}{!}{
			\subfigure[quadratic imputation model]{
				\label{boxplot:a}
				\includegraphics[scale=.5]{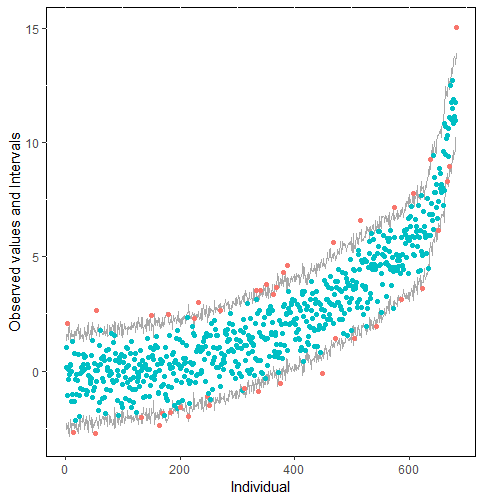}
			}
			\subfigure[linear imputation model]{
				\label{boxplot:b}
				\includegraphics[scale=.5]{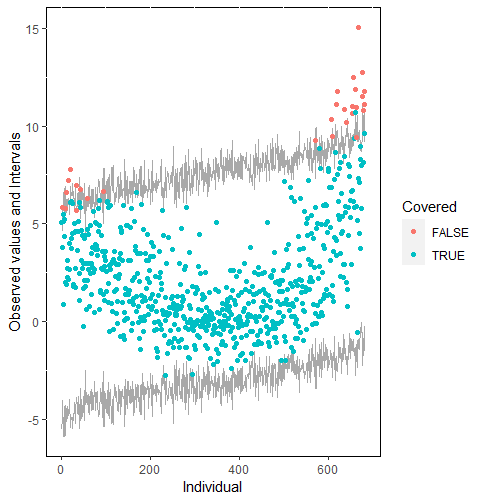}
			}
		}
	\end{center}
	\caption{Distribution plots for the first simulation study (quadratic equation with an incomplete outcome) generated under 30\% missing cases and MARr missingness mechanism. The confidence interval is 95\% nominal. This plot provides upper and lower bounds (grey lines) of the posterior predictive distribution for all observed $Y$ in ascending order of the expectation of the posterior distribution. Blue points imply the corresponding observed value falls in the interval, while red points imply the corresponding observed value falls outside the interval.}
	\label{fig6_2}
\end{figure}

The same result could also be derived from the graphical analysis. Figure \ref{fig6_2} shows distribution plots under the scenario of 30\% missing cases and MARr missing mechanism. This plot provides upper and lower bounds of the posterior predictive distribution for all observed $Y$ in ascending order of the mean of the posterior distribution. Blue points imply the corresponding observed value falls in the interval, while red points imply the corresponding observed value falls outside the interval. 

\begin{figure}[t]
	\begin{center}
		\resizebox{\textwidth}{!}{
			\subfigure[quadratic imputation model]{
				\label{boxplot:a}
				\includegraphics[scale=.5]{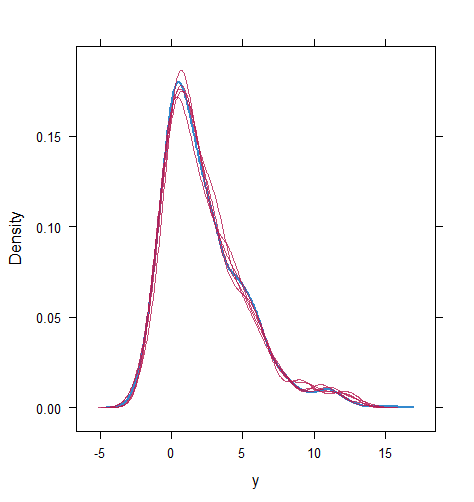}
			}
			\subfigure[linear imputation model]{
				\label{boxplot:b}
				\includegraphics[scale=.5]{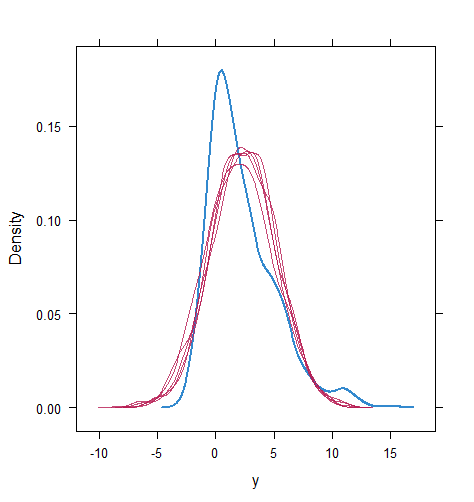}
			}
		}\\ 	
		\resizebox{\textwidth}{!}{
			\subfigure[quadratic imputation model]{
				\label{boxplot:c}
				\includegraphics[scale=.5]{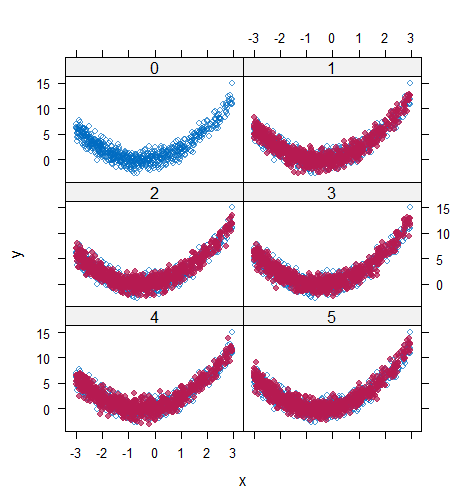}
			}
			\subfigure[linear imputation model]{
				\label{boxplot:d}
				\includegraphics[scale=.5]{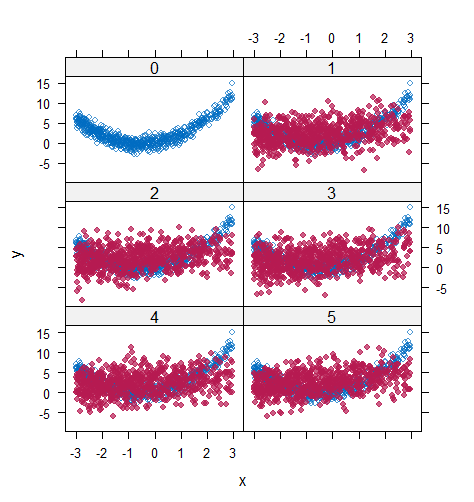}
			}
		}
	\end{center}
	\caption{Scatterplots and densityplots for the first simulation study (quadratic equation with an incomplete outcome) generated under 30\% missing cases and MARr missingness mechanism. Densityplots (a) and (b) show kernel density estimates for the distribution of the variable $Y$ (blue) and $m = 5$ densities calculated from the imputed data (red). Scatterplots (c) and (d) show observed values (blue) of $Y$ (label 0) and $m = 5$ comparisons of observed (blue) and imputed (red) values (label 1-5).}
	\label{fig6_3}
\end{figure}

When the imputation model was correctly specified, the red points were randomly spread over the sample space without any patterns. However, when the imputation model was incorrectly specified as the linear regression model, the red points are shown at tails with lower and higher expectations of the posterior distribution. Moreover, the width of the intervals was generally narrower when the model was correct. The density plot and the scatter plot of the observed and replicated data generated with function \textbf{densityplot()} and \textbf{xyplot()} in \texttt{MICE} also show the evidence that the quadratic regression is preferable than the linear model (see figure \ref{fig6_3}). The scatter plot of the quadratic regression imputation model shows that replicated data overlapped the observed data. The density plot shows that the replicated data shared the same distribution as the observed data. This evidence illustrates the congeniality of the quadratic regression imputation model. However, the linear regression model performed worse than the quadratic regression model. First, the replicated data did not cover the observed data in two extreme regions in the scatter plot. Second, the empirical density of the replicated data and observed data were different. 

\begin{figure}[ht!]
	\begin{center}
		\resizebox{\textwidth}{!}{
			\subfigure[quadratic imputation model]{
				\label{boxplot:a}
				\includegraphics[scale=.5]{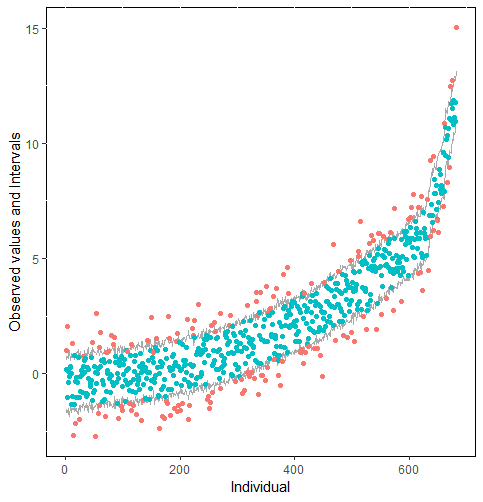}
			}
			\subfigure[linear imputation model]{
				\label{boxplot:b}
				\includegraphics[scale=.5]{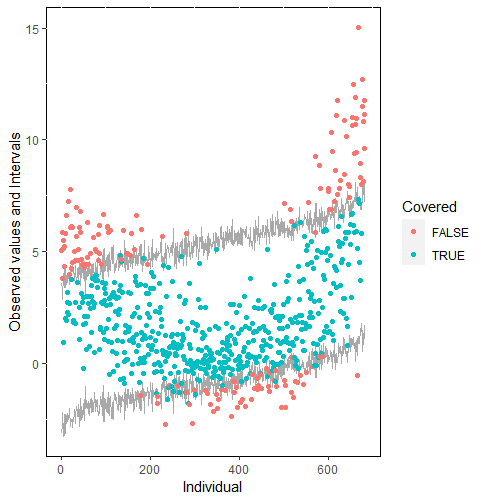}
			}
		}
	\end{center}
	\caption{Distribution plots for the first simulation study (quadratic equation with an incomplete outcome) generated under 30\% missing cases and MARr missingness mechanism. The confidence interval is 75\% nominal. The confidence interval is 95\% nominal. This plot provides upper and lower bounds (grey lines) of the posterior predictive distribution for all observed $Y$ in ascending order of the expectation of the posterior distribution. Blue points imply the corresponding observed value falls in the interval, while red points imply the corresponding observed value falls outside the interval.}
	\label{fig6_4}
\end{figure}

Furthermore, our proposed PPC approach for imputation models is robust against the different percentages of missing cases, missingness mechanisms, and the confidence interval's nominal levels. The nominal level of the confidence interval is determined by the extent to which we could undertake the outliers when the imputation model is not congenial with the data generating process. For instance, there were more outliers in the plot of means and 75\% confidence intervals than the plot of mean and 95\% confidence intervals. (See figure \ref{fig6_4}).

\subsection{Quadratic equation with incomplete covariates}
Tables \ref{tab6_2} and \ref{tab6_3} show the result of the simulation for the quadratic equation with incomplete covariates. Based on the numerical results, the performance of these three methods, PC, SMC-FCS, and PMM, was the same, despite the slightly reduced coverage rate of PMM. In fact, when the missingness mechanism is MCAR (to bypass the problem of the sparse observed region for PMM), PMM would also provide a valid inference of the regression parameters (see Table \ref{tab6_4}) (Vink \& van Buuren, 2013)\nocite{Vink2013}.

\begin{sidewaysfigure}[ht!]
	\begin{center}
		\resizebox{\textwidth}{!}{
			\subfigure[PC]{
				\label{boxplot:a}
				\includegraphics[scale=.5]{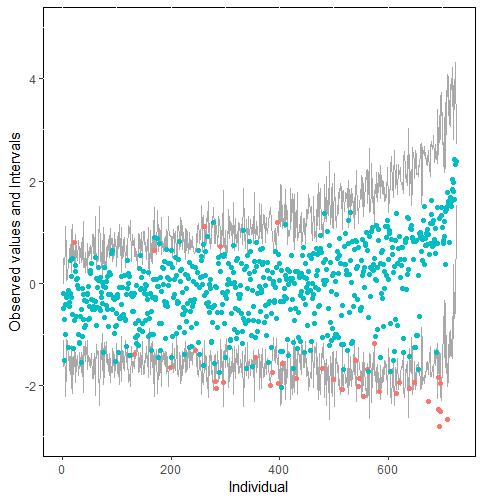}
			}
			\subfigure[SMC-FCS]{
				\label{boxplot:b}
				\includegraphics[scale=.5]{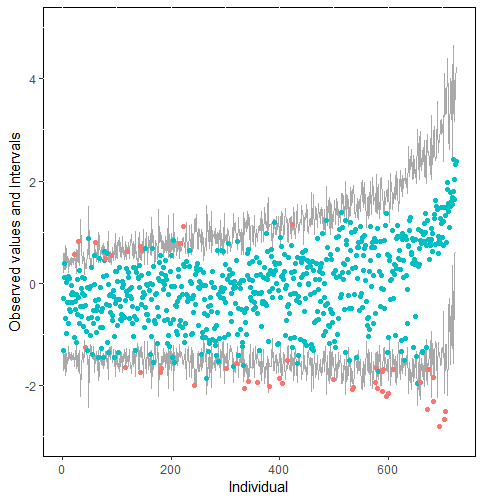}
			}
			\subfigure[PMM]{
				\label{boxplot:b}
				\includegraphics[scale=.5]{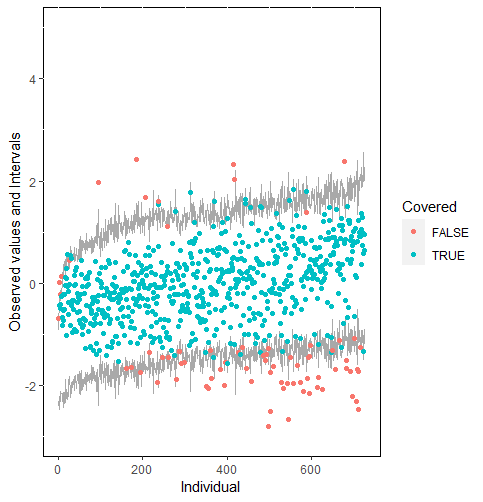}
			}
		}
	\end{center}
	\caption{Distribution plots for the second simulation study (quadratic equation with incomplete covariates) generated under 30\% missing cases and MARr missingness mechanism. The nominal level is 95\%. The confidence interval is 95\% nominal. This plot provides upper and lower bounds (grey lines) of the posterior predictive distribution for all observed $X$ in ascending order of the expectation of the posterior distribution. Blue points imply the corresponding observed value falls in the interval, while red points imply the corresponding observed value falls outside the interval.}
	\label{fig6_5}
\end{sidewaysfigure}

\begin{sidewaysfigure}[ht!]
	\begin{center}
		\resizebox{\textwidth}{!}{
			\subfigure[PC]{
				\label{boxplot:a}
				\includegraphics[scale=.5]{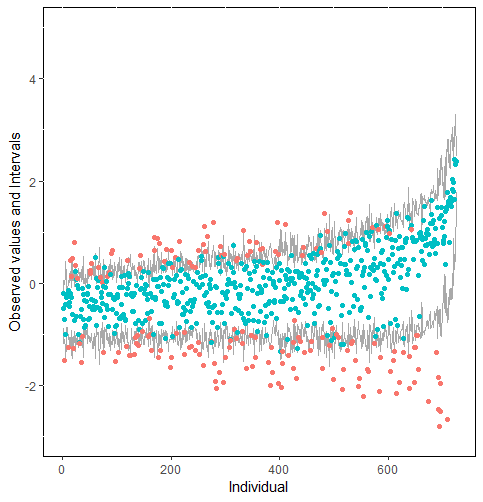}
			}
			\subfigure[SMC-FCS]{
				\label{boxplot:b}
				\includegraphics[scale=.5]{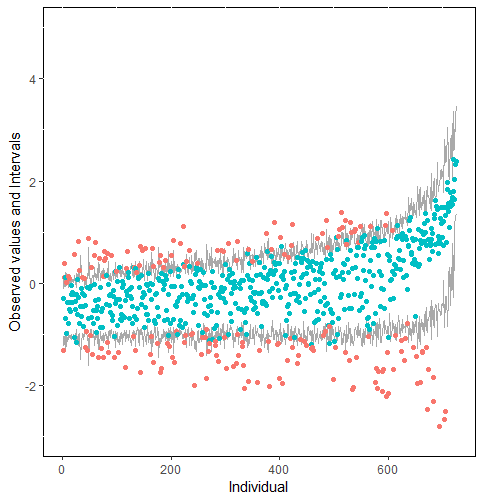}
			}
			\subfigure[PMM]{
				\label{boxplot:b}
				\includegraphics[scale=.5]{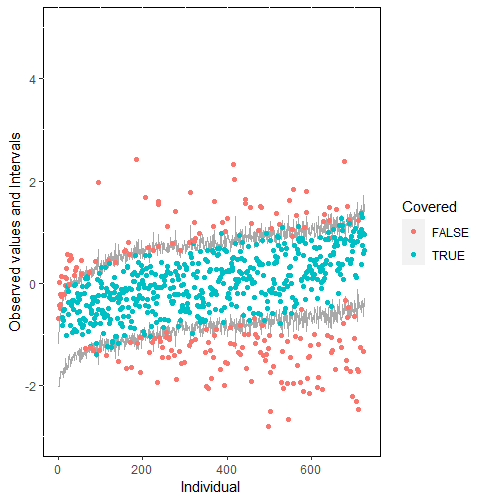}
			}
		}
	\end{center}
	\caption{Distribution plots for the second simulation study (quadratic equation with incomplete covariates) generated under 30\% missing cases and MARr missingness mechanism. The nominal level is 75\%. This plot provides upper and lower bounds (grey lines) of the posterior predictive distribution for all observed $X$ in ascending order of the expectation of the posterior distribution. Blue points imply the corresponding observed value falls in the interval, while red points imply the corresponding observed value falls outside the interval.}
	\label{fig6_6}
	
\end{sidewaysfigure}

\begin{sidewaysfigure}[ht!]
	\begin{center}
		\resizebox{\textwidth}{!}{
			\subfigure[PC]{
				\label{boxplot:a}
				\includegraphics[scale=.5]{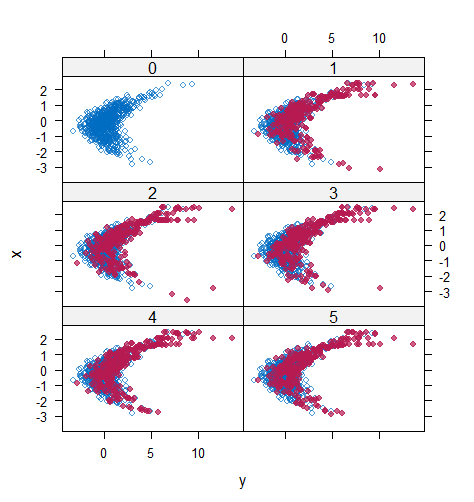}
			}
			\subfigure[SMC-FCS]{
				\label{boxplot:b}
				\includegraphics[scale=.5]{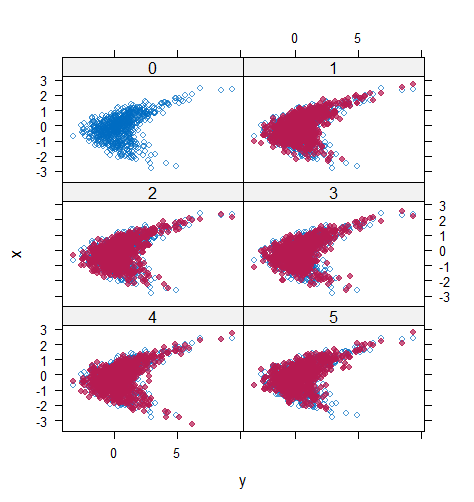}
			}
			\subfigure[PMM]{
				\label{boxplot:b}
				\includegraphics[scale=.5]{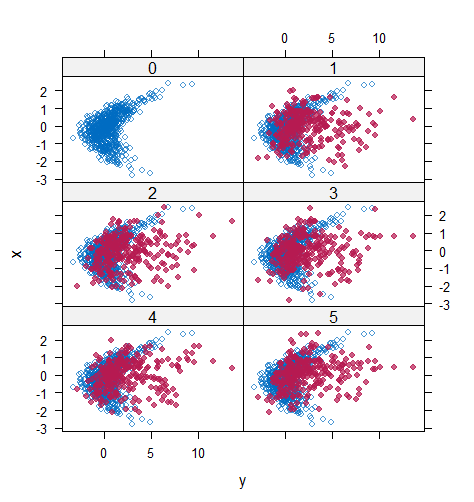}
			}
		}
	\end{center}
	\caption{Scatterplots for the second simulation study (quadratic equation with incomplete covariates) generated under 30\% missing cases and MARr missingness mechanism. Densityplots (a) and (b) show kernel density estimates for the distribution of the variable $X$ (blue) and $m = 5$ densities calculated from the imputed data (red). Scatterplots (c) and (d) show observed values (blue) of $X$ (label 0) and $m = 5$ comparisons of observed (blue) and imputed (red) values (label 1-5).}
	\label{fig6_7}
\end{sidewaysfigure}

However, when it comes to graphical diagnostics, the misfit of PMM appears. The distribution plot (figure \ref{fig6_5} and \ref{fig6_6}) show that PC and SMC-FCS generated the same posterior predictive distribution of the observed data. There were more outliers with a larger value of $Y$. It is sensible since the density function of $X$ based on $Y$ is not monotone. Thus, it is unavoidable to impute the missing cell on the opposite arm of the parabolic function. Although in such a case, the imputed value was not the same as the true value, the replicated data still overlapped the observed data in the scatter plot (see Figures \ref{fig6_7}). The distribution plot of PMM with a 95\% nominal level in Figure \ref{fig6_5} did not show more outliers than these of PC and SMF-FCS. However, when the nominal level was set to 75\%, more outliers appeared in the sub-plot of PMM (Figure \ref{fig6_6}). The reason is that there are more observed data closed to the centre in the plots of PC and SMC-FCS, which implies the superiority of PC and SMC-FCS. The scatter plot also shows the discrepancy between the distribution of the replicated and the observed data with respect to PMM (Figures \ref{fig6_7}). The result is robust against various percentages of missing cases and over the studied missing mechanisms.

\subsection{Generalized linear model for discrete variables}
\begin{figure}[ht!]
	\begin{center}
		\resizebox{\textwidth}{!}{
			\subfigure[logistic model based on $X$ and $Z$]{
				\label{boxplot:a}
				\includegraphics[scale=.5]{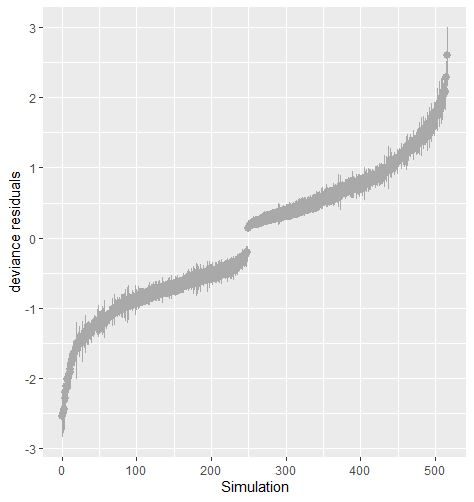}
			}
			\subfigure[logistic model based on $Z$]{
				\label{boxplot:b}
				\includegraphics[scale=.5]{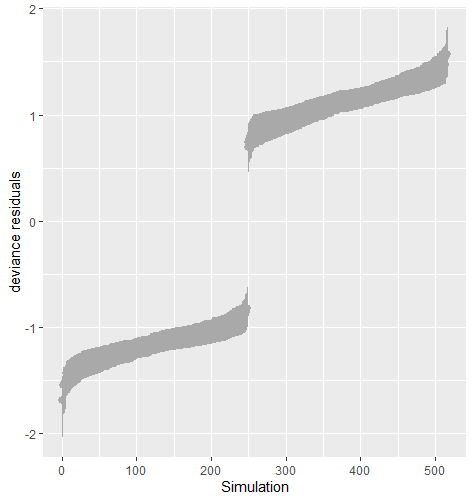}
			}
		}
	\end{center}
	\caption{The plot of deviance residuals for the third simulation study (generalized linear model for discrete variables) generated under two logistic regression imputation models. The percentage of missing is 30\%, and the missingness mechanism is MARr.}
	\label{fig6_8}
\end{figure}

Table \ref{tab6_5} shows the average sum of squared deviance for two different logistic regression models. The value of the average sum of squared deviance was smaller when the imputation model was correctly specified with logistic regression on both $X$ and $Z$. The result is robust against the percentage of missing cases and missingness mechanisms. Figure \ref{fig6_8} shows that the residuals tend to zero when the imputation model fits the observed data better. The distribution of the observed data was more extreme than the empirical posterior distributions of replicated data generated under the logistic model with only variable $Z$.

\section{Application}
\subsection{Background}
We illustrate the application of the proposed PPC for multiple incomplete variables with the data from the body mass index (BMI) of the Dutch adults. This application is to study whether the proposed PPC works for a sequence of imputation models. More specifically, we aim to investigate whether the incorrect imputation model for one incomplete variable would disturb the proposed PPC for other variables. BMI is defined as the body weight divided by the square of the body height, which is broadly applied to categorise a person into underweight, normal, overweight, and obese. Since measuring a person's weight and height is costly, an alternative is to ask people to report their weight and height. However, such self-report data is systematically biased. People are used to overestimating their height and underestimating their weight, leading to a lower self-report BMI compared with measured BMI (van Buuren, 2018, section 9.3). The goal of the study is to estimate unbiased BMI from the self-report data. We apply the multiple imputation approach to fill the unobserved measured weight and height.

The data we analyze is named \texttt{selfreport} in \texttt{MICE} package. The data consists of two components. One is the calibration dataset that contains data on 1257 Dutch individuals with both self-report and measured height and weight which was taken from Krul, Daanen, and Choi (2010)\nocite{krul2011self}. The original survey included 4459 adults from either Italy, Netherlands, or North America aged 18-65 years in 1999 or 2000. The second part is a survey dataset that includes data from 803 Dutch adults aged 18-75 years with only self-reported data. The survey data were collected in November 2007, either online or using paper-and-pencil methods (van Buuren 2018, section 9.3). Six variables are included in the application: \texttt{age} (years), \texttt{sex} (male or female), \texttt{hm} denoting measured height (cm), \texttt{hr} denoting self-reported height (cm), \texttt{wm} denoting measured weight (kg), and \texttt{wr} denoting self-reported weight (kg).

To fit the aim of this application study, we design two linear regression imputation models for \texttt{hm}: one includes all the other variables, and the other includes all the other variables except the variable \texttt{hr}. Similarly, there are two linear regression imputation models for \texttt{wm}: one includes all the other variables, and the other includes all the other variables except the variable \texttt{wr}. In such a case, we have four imputation strategies to evaluate:
\begin{enumerate}
	\item Case 1: include \texttt{hr} in the imputation model of \texttt{hm} and \texttt{wr} in the imputation model of \texttt{wm}.
	\item Case 2: include \texttt{hr} in the imputation model of \texttt{hm} and exclude \texttt{wr} from the imputation model of \texttt{wm}.
	\item Case 3: exclude \texttt{hr} from the imputation model of \texttt{hm} and include \texttt{wr} in the imputation model of \texttt{wm}.
	\item Case 4: exclude \texttt{hr} from the imputation model of \texttt{hm} and \texttt{wr} from the imputation model of \texttt{wm}.
\end{enumerate}
Although it is evident that \texttt{hr} and \texttt{wr} are significant covariates for the imputation of \texttt{hm} and \texttt{wm}, we deliberately ignore these two variables in some cases to evaluate whether incorrect imputation model for \texttt{hm} (\texttt{wm}) influences PPC for \texttt{wm} (\texttt{hm}). If the target of analysis is BMI, one could apply passive imputation to include BMI in the imputation process (van Buuren 2018, section 6.4). In such a case, BMI is still not considered as the predictor of \texttt{hm} and \texttt{hm} because of linear dependencies.  

\subsection{Results}
\begin{figure} [ht!]
	\centering
	\begin{tabular}{cccc}
		\includegraphics[width=0.3\textwidth]{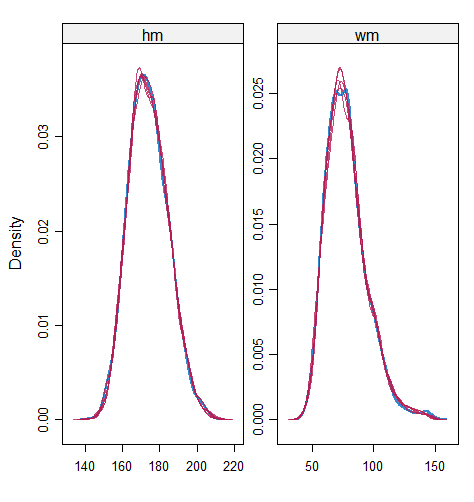} &
		\includegraphics[width=0.3\textwidth]{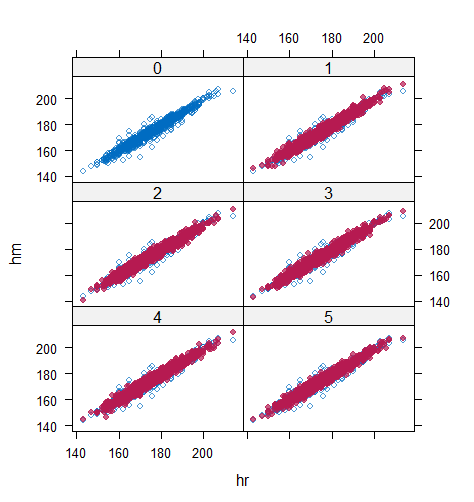} &
		\includegraphics[width=0.3\textwidth]{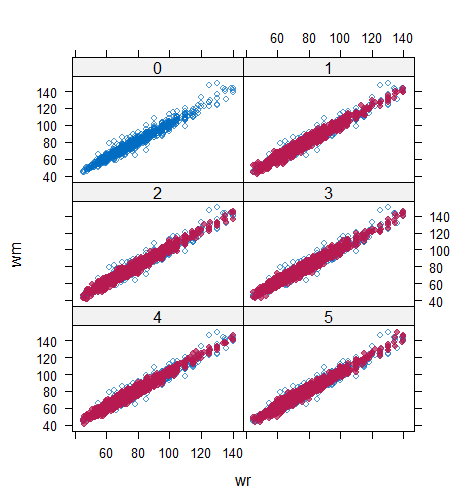} \\
		\textnormal{(a)}  & \textnormal{(b)} & \textnormal{(c)}  \\[6pt]
	\end{tabular}
	\begin{tabular}{cccc}
		\includegraphics[width=0.3\textwidth]{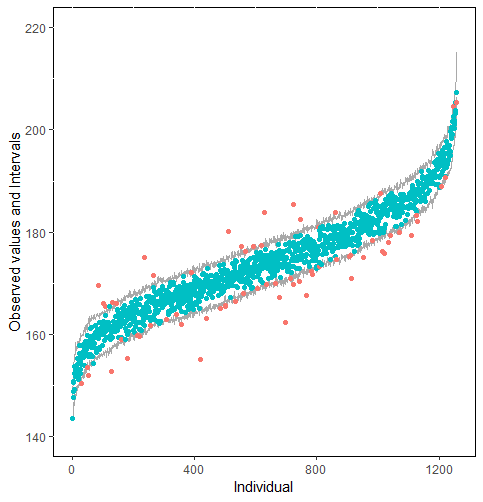} &
		\includegraphics[width=0.3\textwidth]{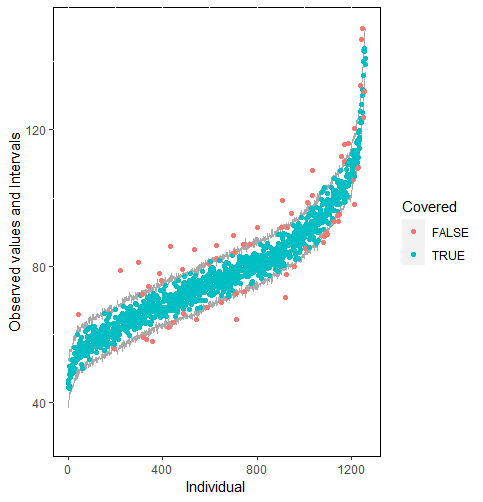} \\
		\textnormal{(d)}  & \textnormal{(e)}  \\[6pt]
	\end{tabular}
	\caption{Graphical analysis of the BMI data with imputation strategy case 1. (a) density plots, (b) scatter plot of \texttt{hm}, (c) scatter plot of \texttt{wm}, (d) distribution plot of \texttt{hm} and (e) distribution plot of \texttt{wm}.}
	\label{fig6_9}
\end{figure}

\begin{figure} [ht!]
	\centering
	\begin{tabular}{cccc}
		\includegraphics[width=0.3\textwidth]{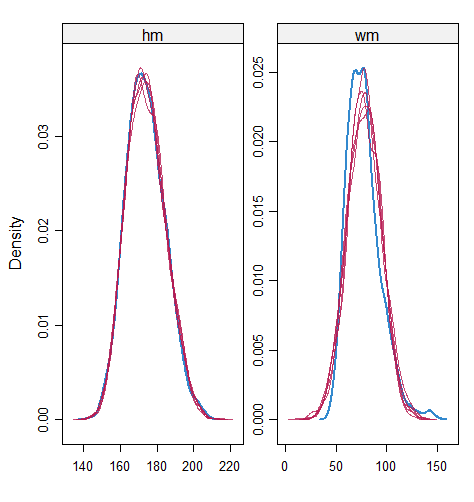} &
		\includegraphics[width=0.3\textwidth]{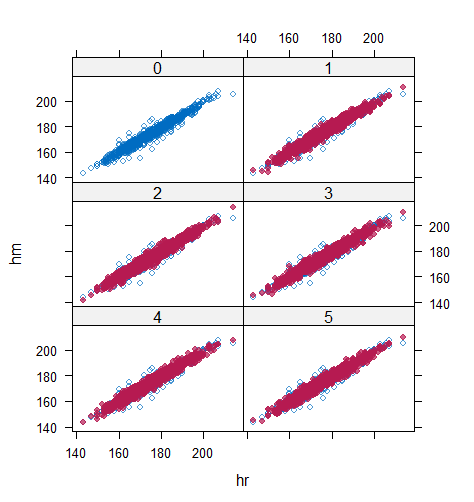} &
		\includegraphics[width=0.3\textwidth]{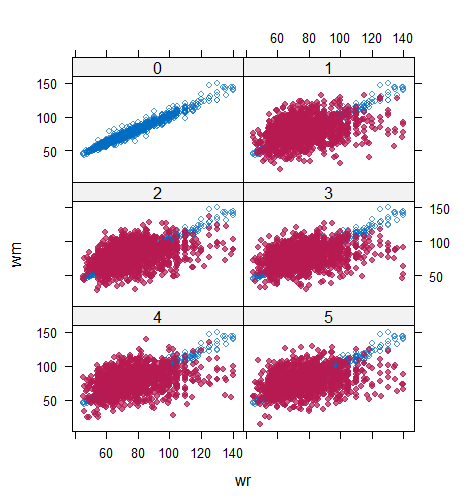} \\
		\textnormal{(a)}  & \textnormal{(b)} & \textnormal{(c)}  \\[6pt]
	\end{tabular}
	\begin{tabular}{cccc}
		\includegraphics[width=0.3\textwidth]{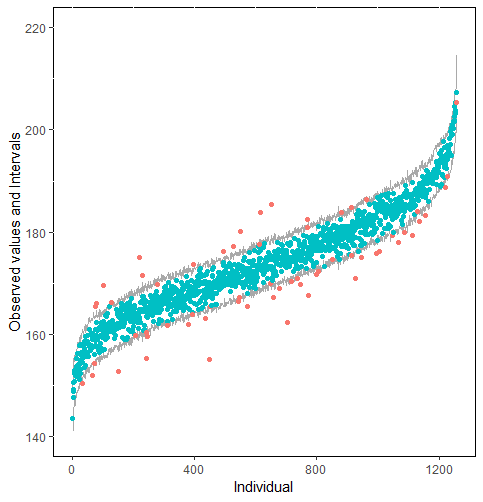} &
		\includegraphics[width=0.3\textwidth]{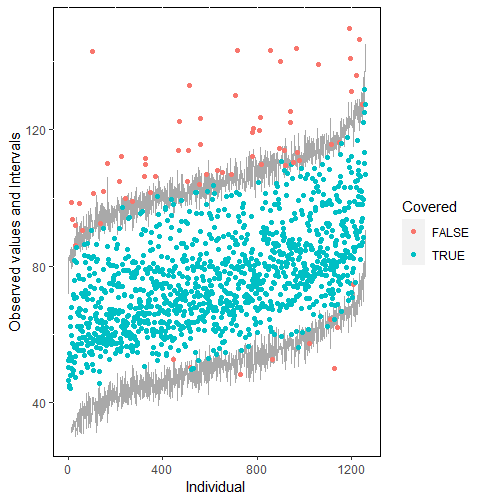} \\
		\textnormal{(d)}  & \textnormal{(e)}  \\[6pt]
	\end{tabular}
	\caption{Graphical analysis of the BMI data with imputation strategy case 2. (a) density plots, (b) scatter plot of \texttt{hm}, (c) scatter plot of \texttt{wm}, (d) distribution plot of \texttt{hm} and (e) distribution plot of \texttt{wm}.}
	\label{fig6_10}
\end{figure}

\begin{figure} [ht!]
	\centering
	\begin{tabular}{cccc}
		\includegraphics[width=0.3\textwidth]{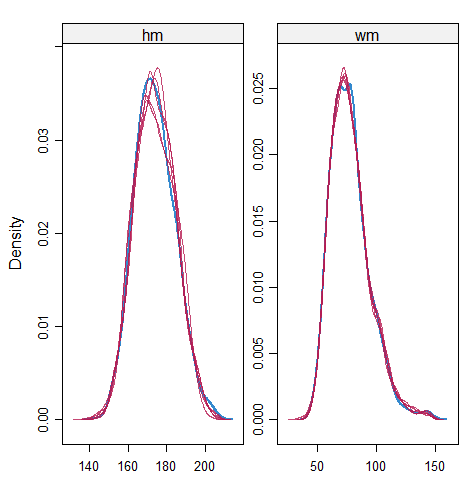} &
		\includegraphics[width=0.3\textwidth]{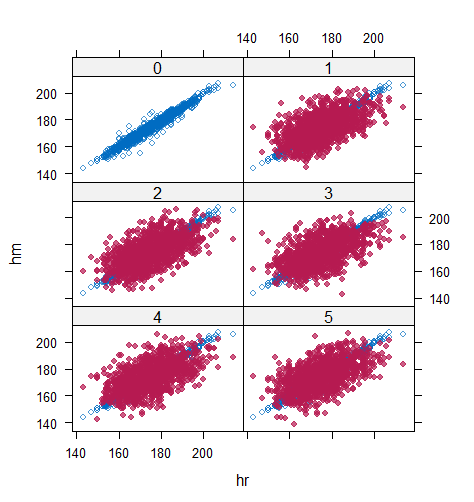} &
		\includegraphics[width=0.3\textwidth]{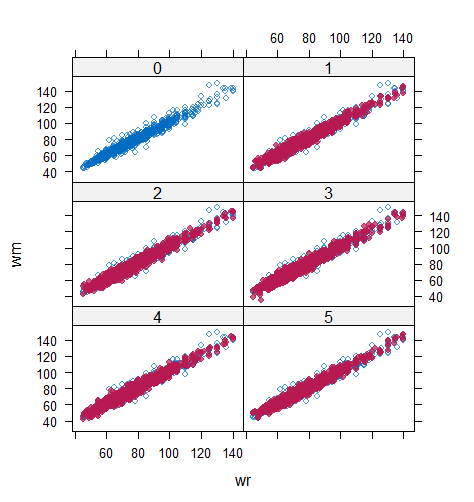} \\
		\textnormal{(a)}  & \textnormal{(b)} & \textnormal{(c)}  \\[6pt]
	\end{tabular}
	\begin{tabular}{cccc}
		\includegraphics[width=0.3\textwidth]{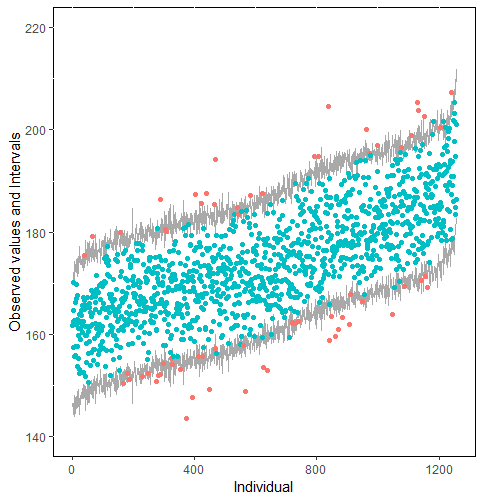} &
		\includegraphics[width=0.3\textwidth]{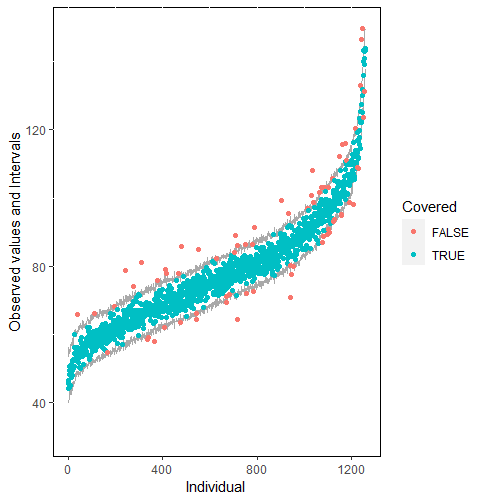} \\
		\textnormal{(d)}  & \textnormal{(e)}  \\[6pt]
	\end{tabular}
	\caption{Graphical analysis of the BMI data with imputation strategy case 3. (a) density plots, (b) scatter plot of \texttt{hm}, (c) scatter plot of \texttt{wm}, (d) distribution plot of \texttt{hm} and (e) distribution plot of \texttt{wm}.}
	\label{fig6_11}
\end{figure}

\begin{figure} [ht!]
	\centering
	\begin{tabular}{cccc}
		\includegraphics[width=0.3\textwidth]{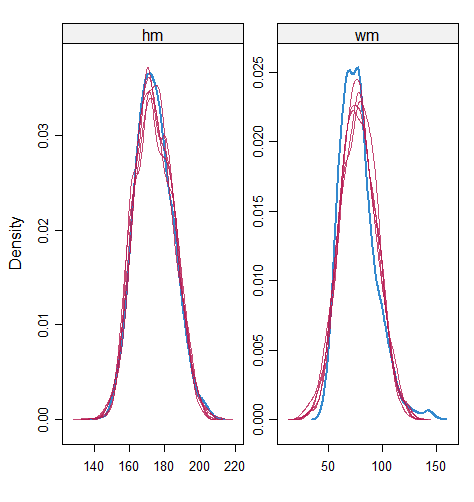} &
		\includegraphics[width=0.3\textwidth]{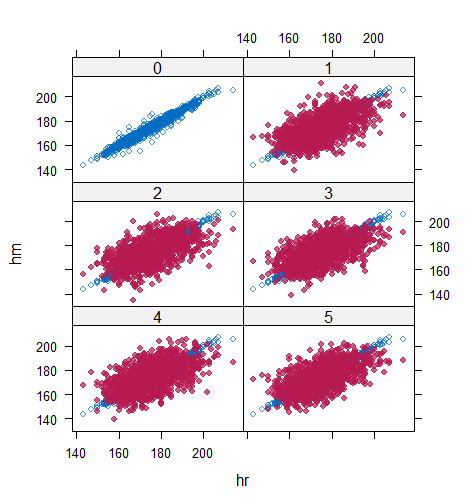} &
		\includegraphics[width=0.3\textwidth]{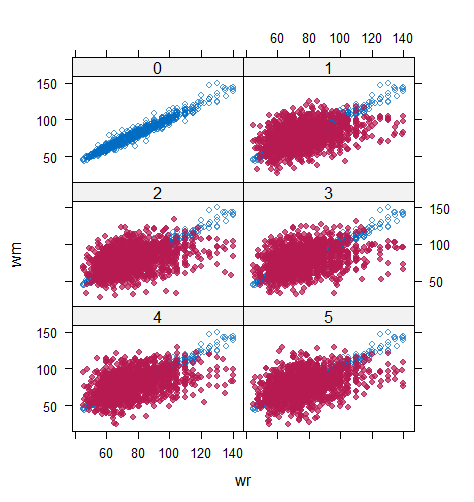} \\
		\textnormal{(a)}  & \textnormal{(b)} & \textnormal{(c)}  \\[6pt]
	\end{tabular}
	\begin{tabular}{cccc}
		\includegraphics[width=0.3\textwidth]{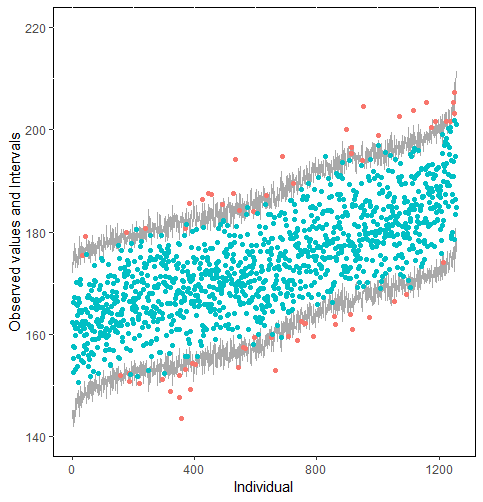} &
		\includegraphics[width=0.3\textwidth]{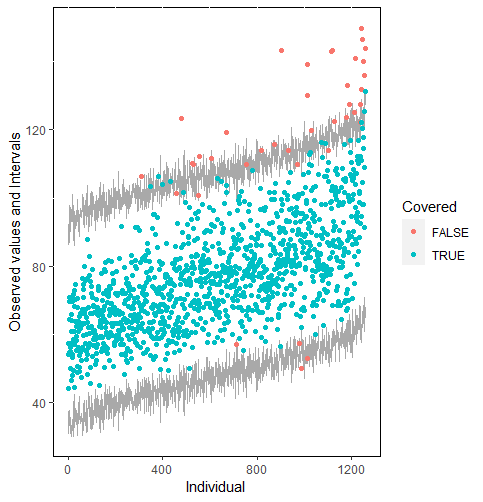} \\
		\textnormal{(d)}  & \textnormal{(e)}  \\[6pt]
	\end{tabular}
	\caption{Graphical analysis of the BMI data with imputation strategy case 4. (a) density plots, (b) scatter plot of \texttt{hm}, (c) scatter plot of \texttt{wm}, (d) distribution plot of \texttt{hm} and (e) distribution plot of \texttt{wm}.}
	\label{fig6_12}
\end{figure}

Table \ref{tab6_6} shows that the best imputation model among these four is the one that includes both \texttt{wr} and \texttt{hr}. The average distance and the width of confidence intervals for the observed data were the smallest for both \texttt{hm} and \texttt{wm}. No matter the imputation model of \texttt{hm} was correctly specified, the linear regression imputation model for \texttt{wm} should be based on all the other variables. When fixing the imputation model for the \texttt{hm} (no matter including \texttt{hr} or not), the average distance and the average width of the confidence interval of \texttt{hm} derived under the linear model included \texttt{hr} was remarkably less than the result taken under the linear model excluded the covariate \texttt{hr}. The graphical results (Figure \ref{fig6_9}-\ref{fig6_12}) show the same conclusion. When the linear regression imputation model for \texttt{wm} or \texttt{hm} was correctly specified, the imputed data overlapped the observed data in the scatter plot. The observed data would be closer to the centre of the confidence interval, and the width of confidence intervals was relatively small. However, the result of \texttt{wr} in case 3 was slightly larger than that in case 1. Similarly, the result of \texttt{wr} in case 4 was slightly larger than that in case 2. A similar result could be found in fixing the imputation model for the \texttt{wm} (no matter the imputation model includes \texttt{wr} or not). The average distance and the average confidence interval of \texttt{wm} derived under the linear model had \texttt{wr} was remarkably less than the result taken under the linear model excluded the covariate \texttt{wr}.

The findings imply that incorrect specification of the imputation models for other incomplete variables $Y_{-j}$ would influence the target variable $Y_j$ for which we perform the PPC because densities of the imputed variables $Y_{-j}$ are different from the `true' densities. However, we can still select the correct model for $Y_j$. Our application scenario is relatively simple: the linear model is sufficient to reflect the data generating process of incomplete variables. However, we do not rule out the possibility that under extreme and complicated cases, incorrect specification of the imputation models for other incomplete variables $Y_{-j}$ would prevent us from selecting the most suitable imputation model for the incomplete variable $Y_j$. 

\section{Discussion}
The proposed imputation model diagnostic procedure based on PPC involves numerical assessment and graphical analysis. It is noteworthy that applying both numerical and graphical tools benefits a thorough understanding of model selection. For numerical assessment, the evidence of a fitted imputation model is less deviation between the observed value and the expectation of corresponding predictive posterior distribution and narrower width of confidence intervals of predictive posterior distributions for the observed data. For graphical analysis, we provide the distribution plot, the scatter plot and the density plot. The more suitable imputation models are, the more similar the replicated data to the observed data in the density and scatter plots. The distribution plot shows posterior distributions of all observed data. It allows the researcher to identify the regions where the imputation model misfits. Furthermore, the graphical analysis could be applied to evaluate whether a specific imputation model is adequate to provide plausible imputations. 

The simulation study demonstrates that the proposed imputation model diagnostic procedure works on continuous and discrete variables under parametric and hot-deck multiple imputation approaches. For continuous variables, the distribution plots can be used to derive information to improve the imputation model. For example, we may identify that although the imputation model is incorrect, it would provide valid imputations in the focused regions. In such a case, we could still apply the suboptimal imputation model. Moreover, we could also adjust the imputation model in the misfitted regions and develop a piecewise imputation model. The PPC for categorical data or ordered categorical data is limited, since the predictor of the imputation model is the probability of assignment rather than the observed data itself. We currently investigate residual deviance as the indicator to select the model for categorical data and ordered categorical data.

For hot-deck imputation approaches, what PPC diagnoses is the donor-selection procedure. As the result shown in section 5.2, selecting donors for the composition $X + X^2$ performed better than only solving for the incomplete variable $X$. SMC-FCS was treated as the baseline in our simulation since it is proven as a reliable imputation method when the substantive model is known (Bartlett et al., 2015)\nocite{bartlett2015multiple}. The PC performs as well as the SMC-FCS which implies the donor selection process of PC reflects the data generating process in our simulation scenarios. However, based on the features of predictive mean matching, the appropriate donor selection does not ensure plausible imputations. Extra analysis of the observed data and the imputed data would then be necessary.

The application example shows that the PPC works on the multivariate incomplete datasets. When the imputed covariate deviates from the actual distribution because of the mis-specified imputation models, the imputation model for the predictor could also be selected by PPC. In our case study, the misspecification of one incomplete variable slightly influences the other incomplete variable's numerical results. However, in more extreme situations, such as a large number of incomplete variables and more ridiculous imputation models for covariates, the result may be influenced seriously, so as to result in a sub-optimal model selection. Therefore, it is more reasonable to perform the numerical analysis of all incomplete variables and make the model selection for those variables with remarkably different results under different candidate imputation approaches first.

Existing PPC proposed by He and Zaslavsky (2012) and Gelman et al. (2005) measured the posterior predictive p-value to indicate the discrepancies of summarised statistics between the observed and replicated data. The close to 0 or 1 p-value implies the inadequacy of the imputation model with respect to the target quantities. The target quantities should be calculated with the completed data, which consists of the observed and the imputed data because it allows the researcher to calculate the target quantities requiring a complete data matrix. Both He and Zaslavsky (2012) and Nguyen et al. (2015) found that the existing PPC for multiple imputation model is sensitive to the percentage of missing cases. Since the imputed and replicated data are generated from the same posterior predictive distribution, evaluation becomes more difficult with an increasing proportion of missing data.

Unlike the existing PPC approach, the PPC discussed in the paper checks the imputation model for each incomplete variable under the FCS framework. We diagnose the distribution of the observed data so that the result would also be reliable with a large proportion of missing cases. The simulation study also shows that the proposed PPC works for different missingness mechanisms.

The PPC for multiple imputation models based on target analysis would be more informative for ``one-goal" studies. The imputer is also the analyst with a specific scientific interest in such a case. The diagnosis procedure aims to select imputation models to produce imputations that could support the particular post-imputation analyses. However, the diagnostic method proposed in this paper is designed for ``multiple-goals" studies. The imputer may not know the potential research on the imputed data. Our approach aims to select congenial imputation models to ensure that Rubin's rule will provide a valid inference. The imputed data generated by selected imputation models could be used for more general downstream analysis and different scientific interests.

When the sample size is tremendous, it is better to choose some representative data to check the imputation model so that the scatter plot or the distribution plot would not be too crowded. A clustered procedure could be applied to gather the observed data with closed values and choose one subset in each cluster to check the model. Further investigation is necessary to set the rule to select the observed data when the sample size is too large.

\newpage
\bibliographystyle{apacite}
\bibliography{PPC}

\begin{thebibliography}{}

\bibitem [\protect \citeauthoryear {%
Abayomi%
, Gelman%
\BCBL {}\ \BBA {} Levy%
}{%
Abayomi%
\ \protect \BOthers {.}}{%
{\protect \APACyear {2008}}%
}]{%
abayomi2008diagnostics}
\APACinsertmetastar {%
abayomi2008diagnostics}%
\begin{APACrefauthors}%
Abayomi, K.%
, Gelman, A.%
\BCBL {}\ \BBA {} Levy, M.%
\end{APACrefauthors}%
\unskip\
\newblock
\APACrefYearMonthDay{2008}{}{}.
\newblock
{\BBOQ}\APACrefatitle {Diagnostics for multivariate imputations} {Diagnostics
  for multivariate imputations}.{\BBCQ}
\newblock
\APACjournalVolNumPages{Journal of the Royal Statistical Society: Series C
  (Applied Statistics)}{57}{3}{273--291}.
\PrintBackRefs{\CurrentBib}

\bibitem [\protect \citeauthoryear {%
Bartlett%
\ \BBA {} Hughes%
}{%
Bartlett%
\ \BBA {} Hughes%
}{%
{\protect \APACyear {2020}}%
}]{%
bartlett2020bootstrap}
\APACinsertmetastar {%
bartlett2020bootstrap}%
\begin{APACrefauthors}%
Bartlett, J\BPBI W.%
\BCBT {}\ \BBA {} Hughes, R\BPBI A.%
\end{APACrefauthors}%
\unskip\
\newblock
\APACrefYearMonthDay{2020}{}{}.
\newblock
{\BBOQ}\APACrefatitle {Bootstrap inference for multiple imputation under
  uncongeniality and misspecification} {Bootstrap inference for multiple
  imputation under uncongeniality and misspecification}.{\BBCQ}
\newblock
\APACjournalVolNumPages{Statistical methods in medical
  research}{29}{12}{3533--3546}.
\PrintBackRefs{\CurrentBib}

\bibitem [\protect \citeauthoryear {%
Bartlett%
, Seaman%
, White%
, Carpenter%
\BCBL {}\ \BBA {} Initiative*%
}{%
Bartlett%
\ \protect \BOthers {.}}{%
{\protect \APACyear {2015}}%
}]{%
bartlett2015multiple}
\APACinsertmetastar {%
bartlett2015multiple}%
\begin{APACrefauthors}%
Bartlett, J\BPBI W.%
, Seaman, S\BPBI R.%
, White, I\BPBI R.%
, Carpenter, J\BPBI R.%
\BCBL {}\ \BBA {} Initiative*, A\BPBI D\BPBI N.%
\end{APACrefauthors}%
\unskip\
\newblock
\APACrefYearMonthDay{2015}{}{}.
\newblock
{\BBOQ}\APACrefatitle {Multiple imputation of covariates by fully conditional
  specification: Accommodating the substantive model} {Multiple imputation of
  covariates by fully conditional specification: Accommodating the substantive
  model}.{\BBCQ}
\newblock
\APACjournalVolNumPages{Statistical methods in medical
  research}{24}{4}{462--487}.
\PrintBackRefs{\CurrentBib}

\bibitem [\protect \citeauthoryear {%
Bondarenko%
\ \BBA {} Raghunathan%
}{%
Bondarenko%
\ \BBA {} Raghunathan%
}{%
{\protect \APACyear {2016}}%
}]{%
bondarenko2016graphical}
\APACinsertmetastar {%
bondarenko2016graphical}%
\begin{APACrefauthors}%
Bondarenko, I.%
\BCBT {}\ \BBA {} Raghunathan, T.%
\end{APACrefauthors}%
\unskip\
\newblock
\APACrefYearMonthDay{2016}{}{}.
\newblock
{\BBOQ}\APACrefatitle {Graphical and numerical diagnostic tools to assess
  suitability of multiple imputations and imputation models} {Graphical and
  numerical diagnostic tools to assess suitability of multiple imputations and
  imputation models}.{\BBCQ}
\newblock
\APACjournalVolNumPages{Statistics in medicine}{35}{17}{3007--3020}.
\PrintBackRefs{\CurrentBib}

\bibitem [\protect \citeauthoryear {%
Cai%
\ \BBA {} Vink%
}{%
Cai%
\ \BBA {} Vink%
}{%
{\protect \APACyear {2022}}%
}]{%
cai2022note}
\APACinsertmetastar {%
cai2022note}%
\begin{APACrefauthors}%
Cai, M.%
\BCBT {}\ \BBA {} Vink, G.%
\end{APACrefauthors}%
\unskip\
\newblock
\APACrefYearMonthDay{2022}{}{}.
\newblock
{\BBOQ}\APACrefatitle {A note on imputing squares via polynomial combination
  approach} {A note on imputing squares via polynomial combination
  approach}.{\BBCQ}
\newblock
\APACjournalVolNumPages{Computational Statistics}{}{}{1--17}.
\PrintBackRefs{\CurrentBib}

\bibitem [\protect \citeauthoryear {%
De~Waal%
, Pannekoek%
\BCBL {}\ \BBA {} Scholtus%
}{%
De~Waal%
\ \protect \BOthers {.}}{%
{\protect \APACyear {2011}}%
}]{%
de2011handbook}
\APACinsertmetastar {%
de2011handbook}%
\begin{APACrefauthors}%
De~Waal, T.%
, Pannekoek, J.%
\BCBL {}\ \BBA {} Scholtus, S.%
\end{APACrefauthors}%
\unskip\
\newblock
\APACrefYear{2011}.
\newblock
\APACrefbtitle {Handbook of statistical data editing and imputation} {Handbook
  of statistical data editing and imputation}\ (\BVOL~563).
\newblock
\APACaddressPublisher{}{John Wiley \& Sons}.
\PrintBackRefs{\CurrentBib}

\bibitem [\protect \citeauthoryear {%
Gelman%
\ \protect \BOthers {.}}{%
Gelman%
\ \protect \BOthers {.}}{%
{\protect \APACyear {2013}}%
}]{%
gelman2013bayesian}
\APACinsertmetastar {%
gelman2013bayesian}%
\begin{APACrefauthors}%
Gelman, A.%
, Carlin, J\BPBI B.%
, Stern, H\BPBI S.%
, Dunson, D\BPBI B.%
, Vehtari, A.%
\BCBL {}\ \BBA {} Rubin, D\BPBI B.%
\end{APACrefauthors}%
\unskip\
\newblock
\APACrefYear{2013}.
\newblock
\APACrefbtitle {Bayesian data analysis} {Bayesian data analysis}.
\newblock
\APACaddressPublisher{}{CRC press}.
\PrintBackRefs{\CurrentBib}

\bibitem [\protect \citeauthoryear {%
Gelman%
, King%
\BCBL {}\ \BBA {} Liu%
}{%
Gelman%
\ \protect \BOthers {.}}{%
{\protect \APACyear {1998}}%
}]{%
gelman1998not}
\APACinsertmetastar {%
gelman1998not}%
\begin{APACrefauthors}%
Gelman, A.%
, King, G.%
\BCBL {}\ \BBA {} Liu, C.%
\end{APACrefauthors}%
\unskip\
\newblock
\APACrefYearMonthDay{1998}{}{}.
\newblock
{\BBOQ}\APACrefatitle {Not asked and not answered: Multiple imputation for
  multiple surveys} {Not asked and not answered: Multiple imputation for
  multiple surveys}.{\BBCQ}
\newblock
\APACjournalVolNumPages{Journal of the American Statistical
  Association}{93}{443}{846--857}.
\PrintBackRefs{\CurrentBib}

\bibitem [\protect \citeauthoryear {%
Gelman%
, Van~Mechelen%
, Verbeke%
, Heitjan%
\BCBL {}\ \BBA {} Meulders%
}{%
Gelman%
\ \protect \BOthers {.}}{%
{\protect \APACyear {2005}}%
}]{%
gelman2005multiple}
\APACinsertmetastar {%
gelman2005multiple}%
\begin{APACrefauthors}%
Gelman, A.%
, Van~Mechelen, I.%
, Verbeke, G.%
, Heitjan, D\BPBI F.%
\BCBL {}\ \BBA {} Meulders, M.%
\end{APACrefauthors}%
\unskip\
\newblock
\APACrefYearMonthDay{2005}{}{}.
\newblock
{\BBOQ}\APACrefatitle {Multiple imputation for model checking: completed-data
  plots with missing and latent data} {Multiple imputation for model checking:
  completed-data plots with missing and latent data}.{\BBCQ}
\newblock
\APACjournalVolNumPages{Biometrics}{61}{1}{74--85}.
\PrintBackRefs{\CurrentBib}

\bibitem [\protect \citeauthoryear {%
He%
\ \BBA {} Zaslavsky%
}{%
He%
\ \BBA {} Zaslavsky%
}{%
{\protect \APACyear {2012}}%
}]{%
he2012diagnosing}
\APACinsertmetastar {%
he2012diagnosing}%
\begin{APACrefauthors}%
He, Y.%
\BCBT {}\ \BBA {} Zaslavsky, A\BPBI M.%
\end{APACrefauthors}%
\unskip\
\newblock
\APACrefYearMonthDay{2012}{}{}.
\newblock
{\BBOQ}\APACrefatitle {Diagnosing imputation models by applying target analyses
  to posterior replicates of completed data} {Diagnosing imputation models by
  applying target analyses to posterior replicates of completed data}.{\BBCQ}
\newblock
\APACjournalVolNumPages{Statistics in medicine}{31}{1}{1--18}.
\PrintBackRefs{\CurrentBib}

\bibitem [\protect \citeauthoryear {%
Heeringa%
}{%
Heeringa%
}{%
{\protect \APACyear {2001}}%
}]{%
heeringa2001multivariate}
\APACinsertmetastar {%
heeringa2001multivariate}%
\begin{APACrefauthors}%
Heeringa, S\BPBI G.%
\end{APACrefauthors}%
\unskip\
\newblock
\APACrefYearMonthDay{2001}{}{}.
\newblock
{\BBOQ}\APACrefatitle {Multivariate imputation of coarsened survey data on
  household wealth.} {Multivariate imputation of coarsened survey data on
  household wealth.}{\BBCQ}
\newblock

\PrintBackRefs{\CurrentBib}

\bibitem [\protect \citeauthoryear {%
Hoogland%
\ \protect \BOthers {.}}{%
Hoogland%
\ \protect \BOthers {.}}{%
{\protect \APACyear {2020}}%
}]{%
hoogland2020handling}
\APACinsertmetastar {%
hoogland2020handling}%
\begin{APACrefauthors}%
Hoogland, J.%
, van Barreveld, M.%
, Debray, T\BPBI P.%
, Reitsma, J\BPBI B.%
, Verstraelen, T\BPBI E.%
, Dijkgraaf, M\BPBI G.%
\BCBL {}\ \BBA {} Zwinderman, A\BPBI H.%
\end{APACrefauthors}%
\unskip\
\newblock
\APACrefYearMonthDay{2020}{}{}.
\newblock
{\BBOQ}\APACrefatitle {Handling missing predictor values when validating and
  applying a prediction model to new patients} {Handling missing predictor
  values when validating and applying a prediction model to new
  patients}.{\BBCQ}
\newblock
\APACjournalVolNumPages{Statistics in medicine}{39}{25}{3591--3607}.
\PrintBackRefs{\CurrentBib}

\bibitem [\protect \citeauthoryear {%
Krul%
, Daanen%
\BCBL {}\ \BBA {} Choi%
}{%
Krul%
\ \protect \BOthers {.}}{%
{\protect \APACyear {2011}}%
}]{%
krul2011self}
\APACinsertmetastar {%
krul2011self}%
\begin{APACrefauthors}%
Krul, A\BPBI J.%
, Daanen, H\BPBI A.%
\BCBL {}\ \BBA {} Choi, H.%
\end{APACrefauthors}%
\unskip\
\newblock
\APACrefYearMonthDay{2011}{}{}.
\newblock
{\BBOQ}\APACrefatitle {Self-reported and measured weight, height and body mass
  index (BMI) in Italy, the Netherlands and North America} {Self-reported and
  measured weight, height and body mass index (bmi) in italy, the netherlands
  and north america}.{\BBCQ}
\newblock
\APACjournalVolNumPages{The European Journal of Public
  Health}{21}{4}{414--419}.
\PrintBackRefs{\CurrentBib}

\bibitem [\protect \citeauthoryear {%
Little%
}{%
Little%
}{%
{\protect \APACyear {1988}}%
}]{%
Little1988}
\APACinsertmetastar {%
Little1988}%
\begin{APACrefauthors}%
Little, R\BPBI J\BPBI A.%
\end{APACrefauthors}%
\unskip\
\newblock
\APACrefYearMonthDay{1988}{jul}{}.
\newblock
{\BBOQ}\APACrefatitle {Missing-Data Adjustments in Large Surveys} {Missing-data
  adjustments in large surveys}.{\BBCQ}
\newblock
\APACjournalVolNumPages{Journal of Business {\&} Economic
  Statistics}{6}{3}{287--296}.
\newblock
\begin{APACrefDOI} \doi{10.1080/07350015.1988.10509663} \end{APACrefDOI}
\PrintBackRefs{\CurrentBib}

\bibitem [\protect \citeauthoryear {%
Meng%
}{%
Meng%
}{%
{\protect \APACyear {1994}}%
}]{%
meng1994multiple}
\APACinsertmetastar {%
meng1994multiple}%
\begin{APACrefauthors}%
Meng, X\BHBI L.%
\end{APACrefauthors}%
\unskip\
\newblock
\APACrefYearMonthDay{1994}{}{}.
\newblock
{\BBOQ}\APACrefatitle {Multiple-imputation inferences with uncongenial sources
  of input} {Multiple-imputation inferences with uncongenial sources of
  input}.{\BBCQ}
\newblock
\APACjournalVolNumPages{Statistical Science}{}{}{538--558}.
\PrintBackRefs{\CurrentBib}

\bibitem [\protect \citeauthoryear {%
Nguyen%
, Carlin%
\BCBL {}\ \BBA {} Lee%
}{%
Nguyen%
\ \protect \BOthers {.}}{%
{\protect \APACyear {2017}}%
}]{%
nguyen2017model}
\APACinsertmetastar {%
nguyen2017model}%
\begin{APACrefauthors}%
Nguyen, C\BPBI D.%
, Carlin, J\BPBI B.%
\BCBL {}\ \BBA {} Lee, K\BPBI J.%
\end{APACrefauthors}%
\unskip\
\newblock
\APACrefYearMonthDay{2017}{}{}.
\newblock
{\BBOQ}\APACrefatitle {Model checking in multiple imputation: an overview and
  case study} {Model checking in multiple imputation: an overview and case
  study}.{\BBCQ}
\newblock
\APACjournalVolNumPages{Emerging themes in epidemiology}{14}{1}{8}.
\PrintBackRefs{\CurrentBib}

\bibitem [\protect \citeauthoryear {%
Nguyen%
, Lee%
\BCBL {}\ \BBA {} Carlin%
}{%
Nguyen%
\ \protect \BOthers {.}}{%
{\protect \APACyear {2015}}%
}]{%
nguyen2015posterior}
\APACinsertmetastar {%
nguyen2015posterior}%
\begin{APACrefauthors}%
Nguyen, C\BPBI D.%
, Lee, K\BPBI J.%
\BCBL {}\ \BBA {} Carlin, J\BPBI B.%
\end{APACrefauthors}%
\unskip\
\newblock
\APACrefYearMonthDay{2015}{}{}.
\newblock
{\BBOQ}\APACrefatitle {Posterior predictive checking of multiple imputation
  models} {Posterior predictive checking of multiple imputation models}.{\BBCQ}
\newblock
\APACjournalVolNumPages{Biometrical Journal}{57}{4}{676--694}.
\PrintBackRefs{\CurrentBib}

\bibitem [\protect \citeauthoryear {%
Rubin%
}{%
Rubin%
}{%
{\protect \APACyear {1987}}%
}]{%
RubinD1987}
\APACinsertmetastar {%
RubinD1987}%
\begin{APACrefauthors}%
Rubin, D\BPBI B.%
\end{APACrefauthors}%
\unskip\
\newblock
\APACrefYear{1987}.
\newblock
\APACrefbtitle {Multiple Imputation for Nonresponse in Surveys} {Multiple
  imputation for nonresponse in surveys}.
\newblock
\APACaddressPublisher{New York}{John Wiley and Sons}.
\PrintBackRefs{\CurrentBib}

\bibitem [\protect \citeauthoryear {%
Schouten%
, Lugtig%
\BCBL {}\ \BBA {} Vink%
}{%
Schouten%
\ \protect \BOthers {.}}{%
{\protect \APACyear {2018}}%
}]{%
Schouten2018}
\APACinsertmetastar {%
Schouten2018}%
\begin{APACrefauthors}%
Schouten, R\BPBI M.%
, Lugtig, P.%
\BCBL {}\ \BBA {} Vink, G.%
\end{APACrefauthors}%
\unskip\
\newblock
\APACrefYearMonthDay{2018}{}{}.
\newblock
{\BBOQ}\APACrefatitle {Generating missing values for simulation purposes: a
  multivariate amputation procedure} {Generating missing values for simulation
  purposes: a multivariate amputation procedure}.{\BBCQ}
\newblock
\APACjournalVolNumPages{Journal of Statistical Computation and
  Simulation}{88}{15}{2909--2930}.
\PrintBackRefs{\CurrentBib}

\bibitem [\protect \citeauthoryear {%
Siddique%
\ \BBA {} Belin%
}{%
Siddique%
\ \BBA {} Belin%
}{%
{\protect \APACyear {2008}}%
}]{%
siddique2008multiple}
\APACinsertmetastar {%
siddique2008multiple}%
\begin{APACrefauthors}%
Siddique, J.%
\BCBT {}\ \BBA {} Belin, T\BPBI R.%
\end{APACrefauthors}%
\unskip\
\newblock
\APACrefYearMonthDay{2008}{}{}.
\newblock
{\BBOQ}\APACrefatitle {Multiple imputation using an iterative hot-deck with
  distance-based donor selection} {Multiple imputation using an iterative
  hot-deck with distance-based donor selection}.{\BBCQ}
\newblock
\APACjournalVolNumPages{Statistics in medicine}{27}{1}{83--102}.
\PrintBackRefs{\CurrentBib}

\bibitem [\protect \citeauthoryear {%
Su%
, Gelman%
, Hill%
\BCBL {}\ \BBA {} Yajima%
}{%
Su%
\ \protect \BOthers {.}}{%
{\protect \APACyear {2011}}%
}]{%
su2011multiple}
\APACinsertmetastar {%
su2011multiple}%
\begin{APACrefauthors}%
Su, Y\BHBI S.%
, Gelman, A\BPBI E.%
, Hill, J.%
\BCBL {}\ \BBA {} Yajima, M.%
\end{APACrefauthors}%
\unskip\
\newblock
\APACrefYearMonthDay{2011}{}{}.
\newblock
{\BBOQ}\APACrefatitle {Multiple imputation with diagnostics (mi) in R: Opening
  windows into the black box} {Multiple imputation with diagnostics (mi) in r:
  Opening windows into the black box}.{\BBCQ}
\newblock

\PrintBackRefs{\CurrentBib}

\bibitem [\protect \citeauthoryear {%
Van~Buuren%
}{%
Van~Buuren%
}{%
{\protect \APACyear {2007}}%
}]{%
van2007multiple}
\APACinsertmetastar {%
van2007multiple}%
\begin{APACrefauthors}%
Van~Buuren, S.%
\end{APACrefauthors}%
\unskip\
\newblock
\APACrefYearMonthDay{2007}{}{}.
\newblock
{\BBOQ}\APACrefatitle {Multiple imputation of discrete and continuous data by
  fully conditional specification} {Multiple imputation of discrete and
  continuous data by fully conditional specification}.{\BBCQ}
\newblock
\APACjournalVolNumPages{Statistical methods in medical
  research}{16}{3}{219--242}.
\PrintBackRefs{\CurrentBib}

\bibitem [\protect \citeauthoryear {%
van Buuren%
}{%
van Buuren%
}{%
{\protect \APACyear {2018}}%
}]{%
Buuren2018}
\APACinsertmetastar {%
Buuren2018}%
\begin{APACrefauthors}%
van Buuren, S.%
\end{APACrefauthors}%
\unskip\
\newblock
\APACrefYear{2018}.
\newblock
\APACrefbtitle {Flexible Imputation of Missing Data, Second Edition} {Flexible
  imputation of missing data, second edition}.
\newblock
\APACaddressPublisher{}{Chapman and Hall/{CRC}}.
\PrintBackRefs{\CurrentBib}

\bibitem [\protect \citeauthoryear {%
Van~Buuren%
\ \BBA {} Groothuis-Oudshoorn%
}{%
Van~Buuren%
\ \BBA {} Groothuis-Oudshoorn%
}{%
{\protect \APACyear {2011}}%
}]{%
van2011mice}
\APACinsertmetastar {%
van2011mice}%
\begin{APACrefauthors}%
Van~Buuren, S.%
\BCBT {}\ \BBA {} Groothuis-Oudshoorn, K.%
\end{APACrefauthors}%
\unskip\
\newblock
\APACrefYearMonthDay{2011}{}{}.
\newblock
{\BBOQ}\APACrefatitle {mice: Multivariate imputation by chained equations in R}
  {mice: Multivariate imputation by chained equations in r}.{\BBCQ}
\newblock
\APACjournalVolNumPages{Journal of statistical software}{45}{1}{1--67}.
\PrintBackRefs{\CurrentBib}

\bibitem [\protect \citeauthoryear {%
Vink%
, Frank%
, Pannekoek%
\BCBL {}\ \BBA {} Van~Buuren%
}{%
Vink%
\ \protect \BOthers {.}}{%
{\protect \APACyear {2014}}%
}]{%
vink2014predictive}
\APACinsertmetastar {%
vink2014predictive}%
\begin{APACrefauthors}%
Vink, G.%
, Frank, L\BPBI E.%
, Pannekoek, J.%
\BCBL {}\ \BBA {} Van~Buuren, S.%
\end{APACrefauthors}%
\unskip\
\newblock
\APACrefYearMonthDay{2014}{}{}.
\newblock
{\BBOQ}\APACrefatitle {Predictive mean matching imputation of semicontinuous
  variables} {Predictive mean matching imputation of semicontinuous
  variables}.{\BBCQ}
\newblock
\APACjournalVolNumPages{Statistica Neerlandica}{68}{1}{61--90}.
\PrintBackRefs{\CurrentBib}

\bibitem [\protect \citeauthoryear {%
Vink%
, Lazendic%
\BCBL {}\ \BBA {} van Buuren%
}{%
Vink%
\ \protect \BOthers {.}}{%
{\protect \APACyear {2015}}%
}]{%
vink2015partioned}
\APACinsertmetastar {%
vink2015partioned}%
\begin{APACrefauthors}%
Vink, G.%
, Lazendic, G.%
\BCBL {}\ \BBA {} van Buuren, S.%
\end{APACrefauthors}%
\unskip\
\newblock
\APACrefYearMonthDay{2015}{}{}.
\newblock
{\BBOQ}\APACrefatitle {Partioned predictive mean matching as a large data
  multilevel imputation technique.} {Partioned predictive mean matching as a
  large data multilevel imputation technique.}{\BBCQ}
\newblock
\APACjournalVolNumPages{Psychological Test and Assessment
  Modeling}{57}{4}{577--594}.
\PrintBackRefs{\CurrentBib}

\bibitem [\protect \citeauthoryear {%
Vink%
\ \BBA {} van Buuren%
}{%
Vink%
\ \BBA {} van Buuren%
}{%
{\protect \APACyear {2013}}%
}]{%
Vink2013}
\APACinsertmetastar {%
Vink2013}%
\begin{APACrefauthors}%
Vink, G.%
\BCBT {}\ \BBA {} van Buuren, S.%
\end{APACrefauthors}%
\unskip\
\newblock
\APACrefYearMonthDay{2013}{sep}{}.
\newblock
{\BBOQ}\APACrefatitle {Multiple Imputation of Squared Terms} {Multiple
  imputation of squared terms}.{\BBCQ}
\newblock
\APACjournalVolNumPages{Sociological Methods {\&} Research}{42}{4}{598--607}.
\newblock
\begin{APACrefDOI} \doi{10.1177/0049124113502943} \end{APACrefDOI}
\PrintBackRefs{\CurrentBib}

\bibitem [\protect \citeauthoryear {%
Volker%
\ \BBA {} Vink%
}{%
Volker%
\ \BBA {} Vink%
}{%
{\protect \APACyear {2021}}%
}]{%
volker2021anonymiced}
\APACinsertmetastar {%
volker2021anonymiced}%
\begin{APACrefauthors}%
Volker, T\BPBI B.%
\BCBT {}\ \BBA {} Vink, G.%
\end{APACrefauthors}%
\unskip\
\newblock
\APACrefYearMonthDay{2021}{}{}.
\newblock
{\BBOQ}\APACrefatitle {Anonymiced Shareable Data: Using mice to Create and
  Analyze Multiply Imputed Synthetic Datasets} {Anonymiced shareable data:
  Using mice to create and analyze multiply imputed synthetic datasets}.{\BBCQ}
\newblock
\APACjournalVolNumPages{Psych}{3}{4}{703--716}.
\PrintBackRefs{\CurrentBib}

\bibitem [\protect \citeauthoryear {%
White%
, Royston%
\BCBL {}\ \BBA {} Wood%
}{%
White%
\ \protect \BOthers {.}}{%
{\protect \APACyear {2011}}%
}]{%
white2011multiple}
\APACinsertmetastar {%
white2011multiple}%
\begin{APACrefauthors}%
White, I\BPBI R.%
, Royston, P.%
\BCBL {}\ \BBA {} Wood, A\BPBI M.%
\end{APACrefauthors}%
\unskip\
\newblock
\APACrefYearMonthDay{2011}{}{}.
\newblock
{\BBOQ}\APACrefatitle {Multiple imputation using chained equations: issues and
  guidance for practice} {Multiple imputation using chained equations: issues
  and guidance for practice}.{\BBCQ}
\newblock
\APACjournalVolNumPages{Statistics in medicine}{30}{4}{377--399}.
\PrintBackRefs{\CurrentBib}

\bibitem [\protect \citeauthoryear {%
Xie%
\ \BBA {} Meng%
}{%
Xie%
\ \BBA {} Meng%
}{%
{\protect \APACyear {2017}}%
}]{%
xie2017dissecting}
\APACinsertmetastar {%
xie2017dissecting}%
\begin{APACrefauthors}%
Xie, X.%
\BCBT {}\ \BBA {} Meng, X\BHBI L.%
\end{APACrefauthors}%
\unskip\
\newblock
\APACrefYearMonthDay{2017}{}{}.
\newblock
{\BBOQ}\APACrefatitle {Dissecting multiple imputation from a multi-phase
  inference perspective: what happens when God's, imputer's and analyst's
  models are uncongenial?} {Dissecting multiple imputation from a multi-phase
  inference perspective: what happens when god's, imputer's and analyst's
  models are uncongenial?}{\BBCQ}
\newblock
\APACjournalVolNumPages{Statistica Sinica}{}{}{1485--1545}.
\PrintBackRefs{\CurrentBib}

\bibitem [\protect \citeauthoryear {%
Yu%
, Burton%
\BCBL {}\ \BBA {} Rivero-Arias%
}{%
Yu%
\ \protect \BOthers {.}}{%
{\protect \APACyear {2007}}%
}]{%
yu2007evaluation}
\APACinsertmetastar {%
yu2007evaluation}%
\begin{APACrefauthors}%
Yu, L\BHBI M.%
, Burton, A.%
\BCBL {}\ \BBA {} Rivero-Arias, O.%
\end{APACrefauthors}%
\unskip\
\newblock
\APACrefYearMonthDay{2007}{}{}.
\newblock
{\BBOQ}\APACrefatitle {Evaluation of software for multiple imputation of
  semi-continuous data} {Evaluation of software for multiple imputation of
  semi-continuous data}.{\BBCQ}
\newblock
\APACjournalVolNumPages{Statistical Methods in Medical
  Research}{16}{3}{243--258}.
\PrintBackRefs{\CurrentBib}

\end{thebibliography}

\newpage
\begin{sidewaystable}[ht!]
	\begin{tabular}{cc|cccc|cccc|cccc}
		\multicolumn{2}{l|}{}                             & \multicolumn{4}{c|}{COV}                                                                                           & \multicolumn{4}{c|}{Average Distance}                                                                              & \multicolumn{4}{c}{Average CIW}                                                                                   \\ \cline{3-14} 
		\multicolumn{1}{l}{}      & \multicolumn{1}{l|}{} & \multicolumn{2}{c}{linear model}                        & \multicolumn{2}{c|}{quadratic model}                     & \multicolumn{2}{c}{linear model}                        & \multicolumn{2}{c|}{quadratic model}                     & \multicolumn{2}{c}{linear model}                        & \multicolumn{2}{c}{quadratic model}                     \\ \hline
		\multicolumn{1}{l|}{}     & \multicolumn{1}{l|}{} & \multicolumn{1}{l}{75\%CI} & \multicolumn{1}{l}{95\%CI} & \multicolumn{1}{l}{75\%CI} & \multicolumn{1}{l|}{95\%CI} & \multicolumn{1}{l}{75\%CI} & \multicolumn{1}{l}{95\%CI} & \multicolumn{1}{l}{75\%CI} & \multicolumn{1}{l|}{95\%CI} & \multicolumn{1}{l}{75\%CI} & \multicolumn{1}{l}{95\%CI} & \multicolumn{1}{l}{75\%CI} & \multicolumn{1}{l}{95\%CI} \\
		\multicolumn{1}{c|}{}     & missingness           &                            &                            &                            &                             &                            &                            &                            &                             &                            &                            &                            &                            \\ \cline{2-14} 
		\multicolumn{1}{c|}{}     & 30                    & 0.76                       & 0.96                       & 0.75                       & 0.93                        & 2.41                       & 2.41                       & 0.79                       & 0.79                        & 6.64                       & 11.32                      & 2.27                       & 3.87                       \\
		\multicolumn{1}{c|}{MCAR} & 50                    & 0.72                       & 0.96                       & 0.76                       & 0.94                        & 2.44                       & 2.44                       & 0.77                       & 0.77                        & 6.61                       & 11.27                      & 2.25                       & 3.83                       \\
		\multicolumn{1}{c|}{}     & 80                    & 0.77                       & 0.95                       & 0.78                       & 0.94                        & 2.25                       & 2.25                       & 0.87                       & 0.87                        & 6.53                       & 11.13                      & 2.51                       & 4.28                       \\\hline
		\multicolumn{1}{c|}{}     & 30                    & 0.75                       & 0.95                       & 0.74                       & 0.94                        & 2.28                       & 2.28                       & 0.8                        & 0.8                         & 6.26                       & 10.66                      & 2.31                       & 3.94                       \\
		\multicolumn{1}{c|}{MARr} & 50                    & 0.76                       & 0.95                       & 0.73                       & 0.95                        & 2.21                       & 2.21                       & 0.81                       & 0.81                        & 6.3                        & 10.73                      & 2.3                        & 3.91                       \\
		\multicolumn{1}{c|}{}     & 80                    & 0.8                        & 0.96                       & 0.77                       & 0.92                        & 1.8                        & 1.8                        & 0.83                       & 0.83                        & 5.43                       & 9.25                       & 2.38                       & 4.05                      
	\end{tabular}
	\caption{The rate of coverage by which the nominal confidence intervals covers the observed data points (COV), the mean of the distance between the observed data and the location of the predictive posterior distribution (Distance), and the average width of the nominal (75\% and 95\%) confidence intervals (CIW) for two imputation models (linear model and quadratic model) under different combinations of experimental factors. The analysis model is a quadratic equation with an incomplete outcome.}
	\label{tab6_1}
\end{sidewaystable}

\begin{sidewaystable}[ht!]
	\begin{tabular}{cc|ccc|ccc|ccc}
		\multicolumn{2}{l}{}                    & \multicolumn{3}{c|}{COV} & \multicolumn{3}{c|}{Average distance} & \multicolumn{3}{c}{Average CIW} \\ \cline{2-11} 
		\multicolumn{1}{c|}{}     & missingness & PC    & SMC-FCS  & PMM   & PC         & SMC-FCS      & PMM       & PC       & SMC-FCS    & PMM     \\
		\multicolumn{1}{c|}{}     & 30          & 0.94  & 0.94     & 0.91  & 0.62       & 0.62         & 0.68      & 3.02     & 3.1        & 3.03    \\
		\multicolumn{1}{c|}{MCAR} & 50          & 0.94  & 0.93     & 0.9   & 0.61       & 0.61         & 0.67      & 3.03     & 3.06       & 2.97    \\
		\multicolumn{1}{c|}{}     & 80          & 0.96  & 0.94     & 0.91  & 0.61       & 0.64         & 0.65      & 3.06     & 3.14       & 3       \\ \hline
		\multicolumn{1}{c|}{}     & 30          & 0.94  & 0.94     & 0.89  & 0.59       & 0.59         & 0.64      & 2.93     & 2.84       & 2.84    \\
		\multicolumn{1}{c|}{MARr} & 50          & 0.93  & 0.94     & 0.89  & 0.57       & 0.58         & 0.62      & 2.74     & 2.7        & 2.68    \\
		\multicolumn{1}{c|}{}     & 80          & 0.93  & 0.97     & 0.93  & 0.56       & 0.56         & 0.61      & 2.61     & 2.64       & 2.69   
	\end{tabular}
	\caption{The rate of coverage by which the nominal confidence intervals covers the observed data points (COV), the mean of the distance between the observed data and the location of the predictive posterior distribution (Distance), and the average width of the nominal 95\% confidence intervals (CIW) for PC, SMC-FCS and PMM under different combinations of experimental factors. The analysis model is a quadratic equation with incomplete covariates.}
	\label{tab6_2}
\end{sidewaystable}

\begin{sidewaystable}[ht!]
	\begin{tabular}{cc|ccc|ccc|ccc}
		\multicolumn{2}{l}{}                    & \multicolumn{3}{c|}{COV} & \multicolumn{3}{c|}{Average distance} & \multicolumn{3}{c}{Average CIW} \\ \cline{2-11} 
		\multicolumn{1}{c|}{}     & missingness & PC    & SMC-FCS  & PMM   & PC         & SMC-FCS      & PMM       & PC       & SMC-FCS    & PMM     \\
		\multicolumn{1}{c|}{}     & 30          & 0.76  & 0.75     & 0.7   & 0.62       & 0.62         & 0.68      & 1.77     & 1.82       & 1.78    \\
		\multicolumn{1}{c|}{MCAR} & 50          & 0.75  & 0.75     & 0.73  & 0.61       & 0.61         & 0.67      & 1.78     & 1.8        & 1.74    \\
		\multicolumn{1}{c|}{}     & 80          & 0.78  & 0.76     & 0.75  & 0.61       & 0.64         & 0.65      & 1.8      & 1.84       & 1.76    \\ \hline
		\multicolumn{1}{c|}{}     & 30          & 0.76  & 0.74     & 0.71  & 0.59       & 0.59         & 0.64      & 1.72     & 1.66       & 1.67    \\
		\multicolumn{1}{c|}{MARr} & 50          & 0.74  & 0.72     & 0.69  & 0.57       & 0.58         & 0.62      & 1.61     & 1.58       & 1.57    \\
		\multicolumn{1}{c|}{}     & 80          & 0.75  & 0.73     & 0.7   & 0.56       & 0.56         & 0.61      & 1.53     & 1.55       & 1.58   
	\end{tabular}
	\caption{The rate of coverage by which the nominal confidence intervals covers the observed data points (COV), the mean of the distance between the observed data and the location of the predictive posterior distribution (Distance), and the average width of the nominal 75\% confidence intervals (CIW) for PC, SMC-FCS and PMM under different combinations of experimental factors. The analysis model is a quadratic equation with incomplete covariates.}
	\label{tab6_3}
\end{sidewaystable}

\begin{table}[ht!]
	\begin{tabular}{cccc}
		& True value & Estimates value & Coverage rate \\
		$\beta_1$ & 1          & 1.008           & 0.934         \\
		$\beta_2$ & 1          & 1               & 0.958        
	\end{tabular}
	\caption{The PMM performs under the scientific model : $Y = \alpha + X\beta_{1} + X^2\beta_{2} +\epsilon$, where $\alpha = 0$, $\beta_{1} = 1$ and $\beta_{2} = 1$. The error term and variable $X$ follow standard normal distributions. 30\% cases of $X$ and $X^2$ are designed to be jointly missing. The missingness mechanism is MCAR.}
	\label{tab6_4}
\end{table}

\begin{table}[ht!]
	\begin{tabular}{cc|cc}
		&             & \multicolumn{2}{c}{mean of residual deviance} \\ \cline{2-4} 
		\multicolumn{1}{c|}{}     & missingness & with x               & without x               \\
		\multicolumn{1}{c|}{}     & 30          & 0.83                 & 1.25                    \\
		\multicolumn{1}{c|}{MCAR} & 50          & 0.85                 & 1.27                    \\
		\multicolumn{1}{c|}{}     & 80          & 0.95                 & 1.3                     \\ \hline
		\multicolumn{1}{c|}{}     & 30          & 0.9                  & 1.34                    \\
		\multicolumn{1}{c|}{MARr} & 50          & 0.94                 & 1.35                    \\
		\multicolumn{1}{c|}{}     & 80          & 0.98                 & 1.28                   
	\end{tabular}
	\caption{The average sum of squared deviance for two imputation models: 1) logistic regression with two predictors $x$ and $z$ 2) logistic regression with one predictor $x$ under different combinations of experimental factors. The outcome is a dichotomous variable $y$ and the binary regression is based on $x$ and $z$.}
	\label{tab6_5}
\end{table}

\begin{table}[ht!]
	\begin{tabular}{c|ccc|ccc}
		& \multicolumn{3}{c|}{hm}               & \multicolumn{3}{c}{wm}                \\ \cline{2-7} 
		& cov  & average distance & average CIW & cov  & average distance & average CIW \\
		strategy 1 & 0.95 & 1.57             & 8.27        & 0.95 & 2.28             & 12.46       \\
		strategy 2 & 0.95 & 1.65             & 8.89        & 0.94 & 10.9             & 54.38       \\
		strategy 3 & 0.95 & 5.58             & 26.89       & 0.94 & 2.35             & 12.83       \\
		strategy 4 & 0.95 & 5.56             & 27.88       & 0.97 & 9.84             & 59.57      
	\end{tabular}
	\caption{The performance of 4 imputation strategies is summarised by the coverage rate, the average distance and the average width of confidence intervals with respect to missing variables \texttt{hm} and \texttt{wm}. \texttt{hm} denotes measured height (cm). \texttt{wm} denotes measured weight (kg). We estimated the rate of coverage by which the nominal confidence intervals covers the observed data points (COV), the mean of the distance between the observed data and the mean of corresponding predictive posterior distributions (Distance), and the average width of the confidence intervals (CIW).}
	\label{tab6_6}
\end{table}

\newpage

\end{document}